\documentclass[aps, twocolumn, groupedaddress, nofootinbib, english, nobalancelastpage, floatfix, superscriptaddress, showpacs]{revtex4-1}

\usepackage{amsmath, amssymb}
\usepackage{graphicx}
\usepackage{epstopdf}
\usepackage{color}
\usepackage{pbox}
\usepackage{slashed}
\usepackage{cancel}

\newcommand{\kpc}{{\rm\; kpc}}
\newcommand{\GeV}{{\rm\; GeV}}
\newcommand{\cm}{{\rm\; cm}}
\newcommand{\s}{{\rm\;s}}
\newcommand{\sv}{\langle\sigma v\rangle}
\newcommand{\svf}{\langle\sigma v\rangle_f}
\newcommand{\Fermi}{{\it Fermi}}
\newcommand{\PV}{\texttt{P6V11}}
\newcommand{\ie}{{\it i.e.~}}
\newcommand{\cf}{{\it cf.~}}

\newcommand{\PPC}{\texttt{PPC4DMID}}

\begin{document}


\title{A Tale of Tails: Dark Matter Interpretations of the \\Fermi GeV Excess
in Light of Background Model Systematics}

\author{Francesca Calore}
\email{f.calore@uva.nl}
\affiliation{GRAPPA, University of Amsterdam, Science Park 904, 1098 XH Amsterdam, Netherlands}

\author{Ilias Cholis}
\email{cholis@fnal.gov}
\affiliation{Center for Particle Astrophysics, Fermi National Accelerator Laboratory, Batavia, IL, 60510, USA}

\author{Christopher McCabe}
\email{c.mccabe@uva.nl}
\affiliation{GRAPPA, University of Amsterdam, Science Park 904, 1098 XH Amsterdam, Netherlands}

\author{Christoph Weniger}
\email{c.weniger@uva.nl}
\affiliation{GRAPPA, University of Amsterdam, Science Park 904, 1098 XH Amsterdam, Netherlands}

\begin{abstract}
    Several groups have identified an extended excess of gamma rays over the
    modeled foreground and background emissions towards the Galactic center
    (GC) based on observations with the \Fermi~Large Area Telescope. This excess
    emission is compatible in morphology and spectrum with a telltale sign from
    dark matter (DM) annihilation. Here, we present a critical reassessment of
    DM interpretations of the GC signal in light of the foreground and
    background uncertainties that some of us recently outlaid in Calore et
    al.~2014.  We find that a much larger number of DM models fits the
    gamma-ray data than previously noted.  In particular: (1) In the case of DM
    annihilation into $\bar{b}b$, we find that even large DM masses up to
    $m_\chi \simeq 74\rm\;GeV$ are allowed at $p$-value $> 0.05$. (2) Surprisingly,
    annihilation into non-relativistic $hh$ gives a good fit to the data.  (3)
    The inverse Compton emission from $\mu^+\mu^-$ with $m_\chi\sim
    60$--$70\rm\;GeV$ can also account for the excess at higher latitudes,
    $|b|>2^\circ$, both in its spectrum and morphology. We also present novel
    constraints on a large number of mixed annihilation channels, including
    cascade annihilation involving hidden sector mediators.  Finally, we show
    that the current limits from dwarf spheroidal observations are not in
    tension with a DM interpretation when uncertainties on the DM halo profile
    are accounted for.
\end{abstract}
\pacs{95.30.Cq,95.35+d,95.85.Pw,FERMILAB-PUB-14-477-A}
\maketitle

\section{Introduction} 

Shedding light onto the origin of Dark Matter (DM) is one of the biggest
challenges of current particle physics and cosmology.  The most appealing
particle DM candidates are the so-called Weakly Interacting Massive Particles
(WIMP) \cite{Jungman:1995df, Bergstrom:2000pn, Bertone:2004pz}.  Among the
different indirect messengers, gamma rays play a dominant role and they have
often been defined as the \emph{golden channel} for DM indirect detection (see
Ref.~\cite{Bringmann:2012ez} for an extensive review).  The main challenge is
to disentangle putative DM signals from the large astrophysical foregrounds and
backgrounds that are generally expected to dominate the measured fluxes.  The
best example of a challenging target is the Galactic Center (GC), where on the
one hand the DM signal is expected to be brighter than anywhere else on the
sky~\cite{Abramowski:2011hc, Hooper:2012sr}, but -- given our poor knowledge of
the conditions in the inner Galaxy -- the astrophysical foreground and
background (either from Galactic point sources or from diffuse emissions) is
subject to very large uncertainties.

In this respect, it is not surprising for unmodeled gamma-ray contributions to
be found towards the inner part of the Galaxy, above or below the expected
standard astrophysical emission.  Indeed, an extended excess in gamma rays at
the GC was reported by different independent groups~\cite{Goodenough:2009gk,
Vitale:2009hr,
Hooper:2010mq, Hooper:2011ti, Abazajian:2012pn, Gordon:2013vta, Macias:2013vya,
Abazajian:2014fta, Daylan:2014rsa, Zhou:2014lva}, using data from the \Fermi\
Large Area Telescope (LAT), and dubbed ``\Fermi~GeV excess'' as it appears to
peak at energies around 1--3 GeV.  Intriguingly, the excess emission shows
spectral and morphological properties consistent with signals expected from DM
particles annihilating in the halo of the Milky Way.  Recently, the existence
of a GeV excess emission towards the GC above the modeled astrophysical
foreground/background was also confirmed by the \Fermi-LAT
Collaboration~\cite{simonaTalk}.  This revitalizes the importance of
understanding the origin of this excess.

Given that the Galactic diffuse emission is maximal along  the Galactic disk
and that a DM signal is expected to be approximately spherical, the preferable
region to search for a DM annihilation signal in \Fermi-LAT data is actually a
region that, depending on the DM profile, extends between a few degrees and a
few tens of degrees away from the GC, above and below the
disk~\cite{Serpico:2008ga, Cholis:2009gv, Bringmann:2011ye,Weniger:2012tx,
Nezri:2012xu, Tavakoli:2013zva}.   Indeed, different groups~\cite{
Hooper:2013rwa, Huang:2013pda, Daylan:2014rsa} extracted an excess with
spectral properties similar to the GeV excess at the GC from the gamma-ray data
at higher Galactic latitudes, up to about $|b|\sim20^\circ$.  The extension to
higher latitudes is a critical test that the DM interpretation had to pass, and
apparently has passed.

\medskip

However, when talking about excesses, a rather central question is: \emph{An
excess above what?} The excess emission is defined above the astrophysical
foregrounds and backgrounds, \ie the Galactic diffuse emission, point sources
and extended sources, modeled in the data analysis.  Most previous studies of
the \textit{Fermi} GeV excess are based on a small number of fixed models for
the Galactic diffuse emission.  These models were built for the sole purpose of
point source analyses and hence introduce uncontrollable systematics in the
analysis of extended diffuse sources.  In addition, since they are the result
of fits to the data, they may falsely absorb part of the putative excess
emission in some of their free components.  This may in turn, lead to biased or
overly constraining statements about the spectrum and morphology of the
\textit{Fermi} GeV excess emission.

To remedy this situation, in Ref.~\cite{Calore:2014xka} some of the present
authors reassessed the spectral and morphological properties of the putative
GeV excess emission from the inner Galaxy\footnote{With \emph{inner Galaxy}
 we refer to the region contained in
few tens of degrees away from the GC and avoiding the very inner few degrees in
latitude. In particular, in Ref.~\cite{Calore:2014xka} the region considered is
$|l| < 20^{\circ}$ and $2^{\circ} < |b| < 20^{\circ}$.}.  Relevant systematic uncertainties
came from the modeling of the Galactic diffuse emission, \Fermi-LAT detected
point sources and the \Fermi~bubbles.  The emission associated with the inner
Galaxy was found to be larger
than expected from standard Galactic diffuse emission models (where the
distribution of the cosmic-ray (CR) sources peaks at kpc distances from the
GC~\cite{FermiLAT:2012aa}).  This ``excess'' is -- by definition -- an excess
above Galactic diffuse emission models that lead to subdominant contribution
from the inner kpc around the GC. It hence should be understood as a
\emph{characterization} of the dominant part of the emission from these central
spatial regions, which is robust w.r.t.~uncertainties in the emission from
other parts along the line-of-sight.  This emission features a spectrum that
rises at energies below 1 GeV with a spectral index harder than two, peaks at
1--3 GeV, and has a high-energy tail that continues up to 100 GeV.  The large
uncertainties in the spectrum were estimated from a study of residuals along
the Galactic disk.  The observed emission was found to be consistent with the
hypothesis of a uniform gamma-ray energy spectrum at 95\% CL, with spherical
symmetry around the GC and a radial extension of at least 1.5~kpc.

The proper treatment of systematic uncertainties has important consequences for
the interpretation of the \textit{Fermi} GeV excess: Although all studies find
that the emission peaks around 1--3 GeV, the \emph{low- and high-energy tails}
of the spectrum are much more uncertain.  As we will see below, this allows
significant -- previously ignored -- freedom for DM models fitting the excess.

\medskip

In what follows, we focus on the DM interpretations of the excess emission.  We
will therefore not further discuss potential astrophysical explanations, like
the emission from an unresolved population of point-like sources concentrated
in the very center of the Galaxy (see~\cite{Hooper:2013nhl, Calore:2014oga,
Cholis:2014lta, Petrovic:2014xra} for relevant discussions) or the injection of
leptons and/or protons during a burst event at the GC some kilo-/mega-years
ago~\cite{Carlson:2014cwa, Petrovic:2014uda}.  In particular, we will here
entertain the possibility that \emph{all} of the excess emission is coming from
a single diffuse source.  This obviously does not have to be the case, but it
is a suggestive (and from the perspective of a particle physicist minimal)
assumption, given the uniform spectral properties of the emission in different
regions of the sky~\cite{Calore:2014xka}.

As for the DM interpretation, there is by now an extensive array of DM
scenarios that both can explain the observed emission by DM annihilation while
simultaneously being compatible with other direct, indirect and collider
constraints~\cite{Logan:2010nw, Buckley:2010ve, Zhu:2011dz, Marshall:2011mm,
Boucenna:2011hy, Buckley:2011mm, Anchordoqui:2013pta, Buckley:2013sca,
Hagiwara:2013qya, Okada:2013bna, Huang:2013apa, Modak:2013jya, Boehm:2014hva,
Alves:2014yha,Berlin:2014tja,Agrawal:2014una,Izaguirre:2014vva,
Cerdeno:2014cda, Ipek:2014gua,Boehm:2014bia,Ko:2014gha, Abdullah:2014lla,
Ghosh:2014pwa, Martin:2014sxa, Basak:2014sza, Berlin:2014pya, Cline:2014dwa,
Han:2014nba, Detmold:2014qqa, Wang:2014elb, Chang:2014lxa, Arina:2014yna,
Cheung:2014lqa, McDermott:2014rqa, Huang:2014cla,
Balazs:2014jla,Ko:2014loa,Okada:2014usa,Ghorbani:2014qpa,
Banik:2014eda,Borah:2014ska,Cahill-Rowley:2014ora,Guo:2014gra,Freytsis:2014sua,
Heikinheimo:2014xza, Arcadi:2014lta, Richard:2014vfa, Bell:2014xta}.
The most relevant \emph{indirect} constraints on DM models come from the
non-observation of spectral features in the AMS-02 measurements of CR
 positrons~\cite{Bergstrom:2013jra, Ibarra:2013zia}, and PAMELA
observations of the CR anti-protons~\cite{Bringmann:2014lpa, Cirelli:2014lwa,
Evoli:2011id, Cholis:2010xb, Donato:2008jk, Kappl:2011jw} (see however
Ref.~\cite{Hooper:2014ysa}).

\medskip

Another important set of targets for indirect DM searches, which are, by
comparison to the GC, more simple targets, are \emph{dwarf spheroidal galaxies}
(dSph).  No gamma-ray emission has been detected so far from such objects, and
strong constraints have been set on the DM annihilation
signals~\cite{Geringer-Sameth:2014qqa, Cholis:2012am, GeringerSameth:2011iw,
Abdo:2010ex}.  These results are in general considered to be rather robust (see
however Ref.~\cite{Cholis:2012am}), and we will discuss in detail the impact of
these limits on the DM interpretation of the \Fermi\ GeV excess.

\bigskip

The goal of the present paper is three-fold:  First, we will characterize, for
the first time and in a coherent way, the impact of foreground model
systematics as discussed in Ref.~\cite{Calore:2014xka} on possible DM
interpretations of the \textit{Fermi} GeV excess, and show that a much larger
number of DM models is viable than what was claimed before.  Second, we will
elaborate on the role of Inverse Compton Scattering (ICS) emission at higher
latitudes in the case of leptonic channels.  And third, we will discuss the
impact of recent limits from dSphs on the DM interpretation of the
\textit{Fermi} GeV excess.

\smallskip

The paper is organized as follows:  In Sec.~\ref{sec:bkg}, we discuss the
(non-)consistency of previous results for the intensity of the \textit{Fermi}
GeV excess at energies of 2 GeV, with emphasis on the higher-latitude
\emph{tail} of the emission.  In Sec.~\ref{sec:fitting}, we revise the main
contributions to the gamma-ray sky coming from the CR interactions with the
interstellar medium and we summarize the uncertainties affecting the modeling
of the Galactic diffuse emission.  We then describe how these uncertainties
affect the low- and high-energy \emph{tails} of the energy spectrum of the
\textit{Fermi} GeV excess.  In Sec.~\ref{sec:results}, we discuss possible
models for the DM interpretation of the \textit{Fermi} GeV excess by analyzing
different pure and mixed final states with and without inverse Compton
emission.  Last but not least, in Sec.~\ref{sec:dwarfs}, we compare the
findings about the \textit{Fermi} GeV excess with the current constraints on
the DM parameter space coming from the analysis of dwarf spheroidal galaxies,
in light of observational constraints on the DM halo of the Milky Way.  In
Sec.~\ref{sec:conclusions} we conclude.

\section{The ``\textit{Fermi} GeV excess'' as a genuine feature in the
gamma-ray sky}
\label{sec:bkg} 

\begin{figure*}
    \begin{center}
        \includegraphics[width=0.9\linewidth]{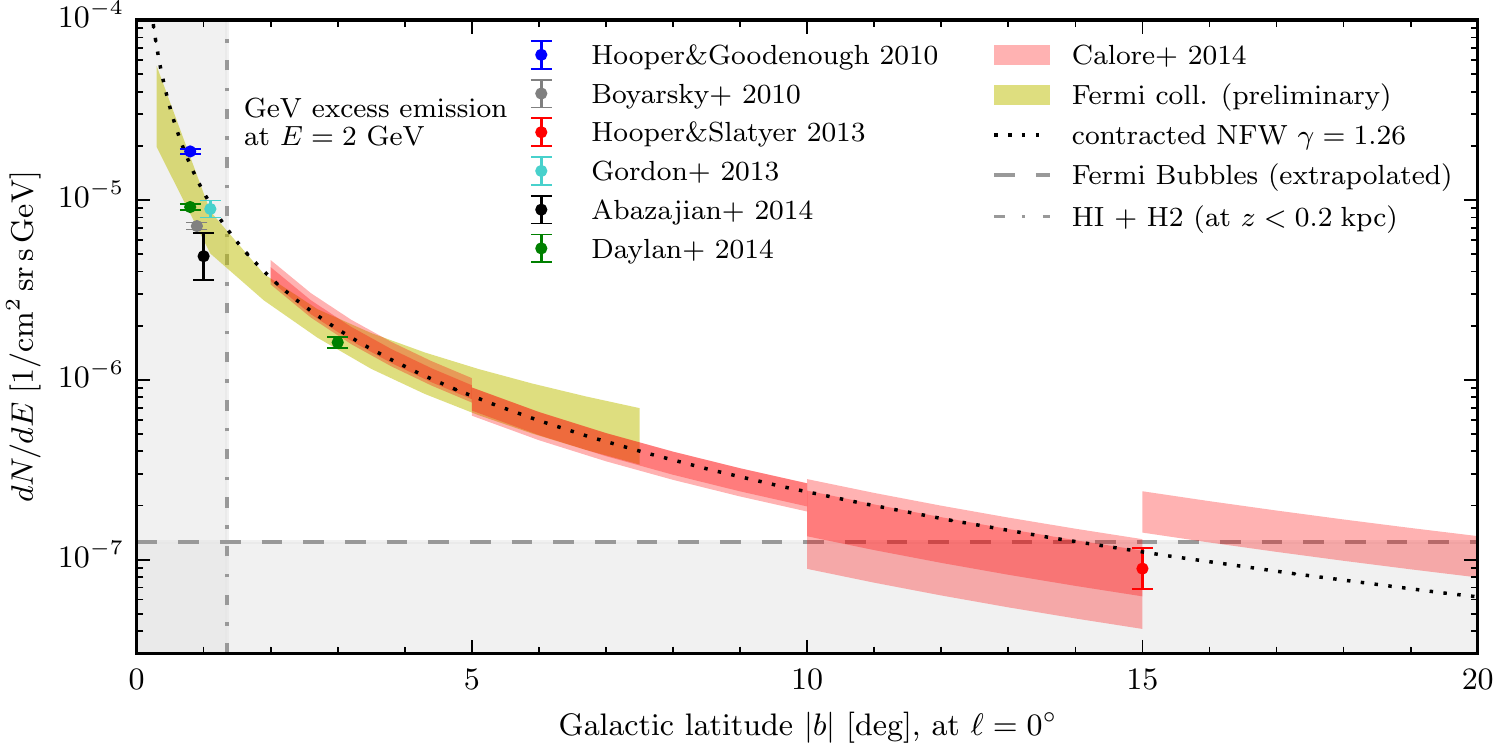}
    \end{center}
    \caption{Intensity of the \textit{Fermi} GeV excess at 2 GeV as function of
    Galactic latitude (see text for details), compared with the expectations
    for a contracted NFW profile (\emph{dotted line}).  Error bars refer to
    statistical $\pm1\sigma$ uncertainties, except for
    Refs.~\cite{Macias:2013vya, Abazajian:2014fta} for which we take into
    account the quoted systematics coming from different astrophysical models.
    The result from Ref.~\cite{Calore:2014xka} for the \emph{higher-latitude
    tail} and the preliminary results by the \Fermi-LAT team~\cite{simonaTalk}
    on the Galactic center include an estimate of the impact of foreground
    systematics.  In these cases, the adopted ROIs are shown as bands (for
    Ref.~\cite{Calore:2014xka}, overlapping regions correspond to the north and
    south parts of the sky).  \emph{Gray areas} indicate the intensity level of
    the \Fermi~bubbles, extrapolated from $|b|>10^\circ$, and the region where
    HI and H2 gas emission from the inner Galaxy becomes important.}
    \label{fig:measured_profile}
\end{figure*}

In Fig.~\ref{fig:measured_profile} we present a convenient comparison of the
differential intensity of the \textit{Fermi} GeV excess emission as derived by
different groups, both for the GC in the inner few degrees, as well as the
\emph{higher-latitude tail} up to $\psi\sim 20^\circ$.  We show the
differential intensity at a reference energy of 2 GeV. At this energy the
putative excess emission is -- compared to other foregrounds/backgrounds --
strongest, so the uncertainties due to foreground/background subtraction
systematics are expected to be the smallest.

The intensities were derived by a careful rescaling of results in the
literature that fully takes into account the assumed excess profiles. In most
works, intensities are quoted as averaged over a given Region Of Interest
(ROI).  Instead of showing these averaged values, which depend on the details
of the adopted ROI, we use the excess profiles to calculate the differential
intensity at a fixed angular distance from the GC.  These excess profiles
usually follow the predictions similar to those of a DM annihilation profile
from a generalized Navarro-Frenk-White (NFW) density distribution, which is
given by
\begin{equation}
    \rho(r) = \rho_s \frac{r_s^3}{r^\gamma(r+r_s)^{3-\gamma}}\;.
\end{equation}
Here, $r_s$ denotes the scale radius, $\gamma$ the slope of the inner part of
the profile, and $\rho_s$ the scale density.  As reference values we will -- if
not stated otherwise -- adopt $r_s=20\kpc$ and $\gamma=1.26$, and $\rho_s$ is
fixed by the requirement that the local DM density at $r_\odot = 8.5\kpc$ is
$\rho_\odot = 0.4\GeV\cm^{-3}$~\cite{Catena:2009mf, Salucci:2010qr}.   

We note that the intensities that we quote from Ref.~\cite{Calore:2014xka}
refer already to a $\bar{b}b$ spectrum and take into account correlated
foreground systematics as discussed below.  
In Ref.~\cite{Calore:2014xka} a broken power-law was found to give a fit as good as 
the DM $\bar{b}b$  spectrum. Assuming a broken power-law, the intensities in Fig.~\ref{fig:measured_profile} 
would be somewhat larger.

\medskip

We find that all previous and current results (with the exception of
Ref.~\cite{Goodenough:2009gk}, which we do not show in
Fig.~\ref{fig:measured_profile}) agree within a factor of about two with a
signal morphology that is compatible with a contracted NFW profile with slope
$\gamma=1.26$, as it was noted previously~\cite{Daylan:2014rsa,
Calore:2014xka}.  As mentioned in our Introduction, the indications for a
higher-latitude tail of the GeV excess profile is a rather non-trivial test for
the DM interpretation and provides a serious benchmark for any astrophysical
explanation of the excess emission.  However, we have to caution that most of
the previous analyses make use of the \emph{same} model for Galactic diffuse
emission (\texttt{P6V11}).  An agreement between the various results is hence
not too surprising.  Instead in the work of Ref.~\cite{Calore:2014xka}, the
$\pi^{0}$, bremsstrahlung and ICS emission maps, where calculated as
independent components, with their exact morphologies and spectra as predicted
from a wide variety of foreground/background models.  As it was shown in
Ref.~\cite{Calore:2014xka}, the exact assumptions on the CR propagation and the
Galactic properties along the line-of-sight can impact both the spectrum and
the morphology (which also vary with energy) of the individual gamma-ray
emission maps.  To probe the associated uncertainties on those diffuse
emissions, the authors of Ref.~\cite{Calore:2014xka} built different models allowing
for extreme assumptions on the CR sources distribution and injection spectra,
on the Galactic gasses distributions, on the interstellar radiation field
properties, on the Galactic magnetic field magnitude and profile and on the
Galactic diffusion, convection and re-acceleration.  

Having performed these tests, it is reassuring that Ref.~\cite{Calore:2014xka}
and later on Ref.~\cite{simonaTalk}, which employs an independently derived array of
foreground/background models, find -- in their respective ROIs and around 2 GeV
-- results that agree both in morphology and intensity of the \textit{Fermi}
GeV excess emission, between themselves and with previous
works.\footnote{Although the intensity of the \Fermi~GeV excess that was found
in Ref.~\cite{simonaTalk} agrees at 2 GeV with previous findings, one has to be
careful with using the preliminary energy spectra presented in that work for
spectral studies.  In particular for two of the presented background models,
the spectral slopes of the background components were explicitly not
tuned to match the observations.  This may bias the residual gamma-ray excess
towards higher energies, which could lead to biased results when fitting the excess
spectrum.}

In Fig.~\ref{fig:measured_profile}, we also indicate the latitude regions where
the flux from the \Fermi\ bubbles becomes important (at $|b|\gtrsim 14^\circ$,
assuming a uniform intensity extrapolated from higher latitudes) and where
strong emission from HI+H2 gas in the inner Galaxy might significantly affect
the results (the inner 0.2 kpc).  It appears that the latitude range
$2^\circ\leq|b|\leq 15^\circ$ is best suited to extract spectral information
about the GeV excess.

Despite the agreement, from Fig.~\ref{fig:measured_profile} it is also evident
that the exact values of the intensities disagree with each other at the
$>3\sigma$ level. Since most of the error bars are statistical only, this
confirms that systematic uncertainties in the subtraction of diffuse and point
source emission play a crucial role for the excess intensity.  \emph{These
effects will be even more important for the spectral shape of the excess.}  We
will concentrate on the implication of Galactic diffuse model systematics for
DM models in the next two sections~\ref{sec:fitting} and~\ref{sec:results}.

\section{The tails in the \textit{Fermi} GeV excess spectrum}
\label{sec:fitting}

\begin{figure}
    \begin{center}
        \includegraphics[width=0.9\linewidth]{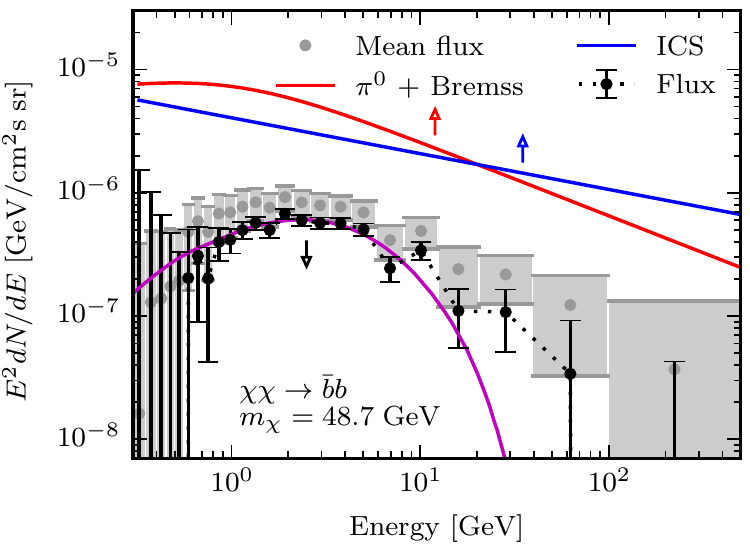}
        \includegraphics[width=0.9\linewidth]{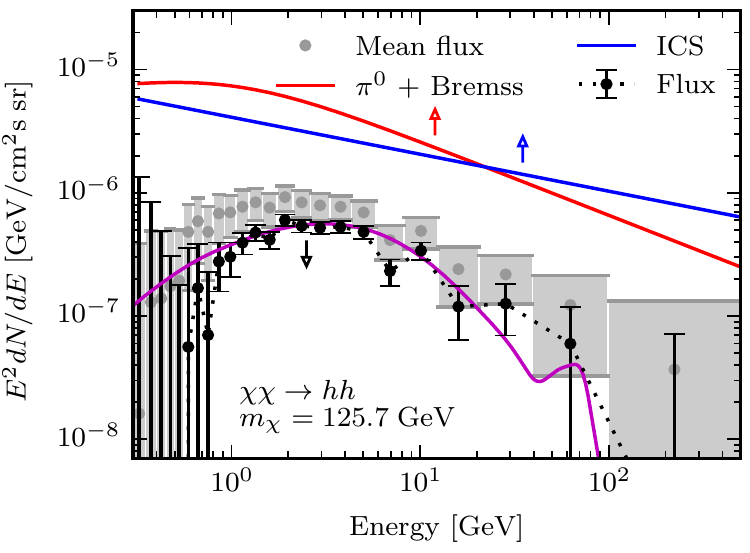}
        \includegraphics[width=0.9\linewidth]{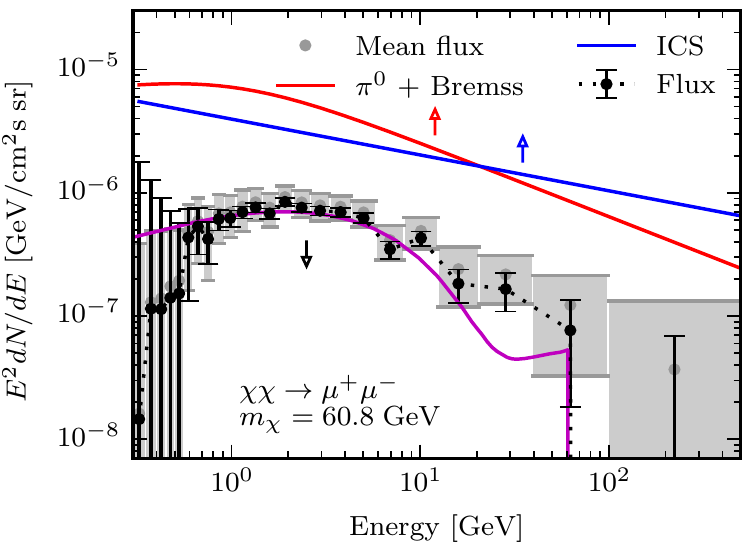}
    \end{center}
    \caption{The foreground/background systematics as derived in
    Ref.~\cite{Calore:2014xka} allow a large number of DM annihilation channels
    to fit the data.  This is here illustrated for three best-fit channels from
    Tabs.~\ref{tab:fitResults} and~\ref{tab:fitResultsICS} (taking model F).
    Correlated systematics are shown by the \emph{gray bands}, uncorrelated
    statistical errors by the error bars (including also remaining method
    uncertainties~\cite{Calore:2014xka}), and we show the estimated ICS and
    $\pi^0$+Bremss foreground/background fluxes for comparison.  As illustrated
    by the \emph{black dots}, a small \emph{increase} of these estimated Galactic
    diffuse emissions within their systematic uncertainties (barely visible on
    the log-scale) leads to a \emph{decrease} of the inferred \textit{Fermi}
    GeV excess flux and vice-versa.  The magnitude of this effect is dependent
    on the fitted spectrum (and hence different in the three panels), but
    automatically taken into account when the full covariance matrix is used.
    Fluxes are averaged over $|l| < 20^{\circ}$ 
    and $2^{\circ} < |b| < 20^{\circ}$.}
    \label{fig:flux}
\end{figure}

As already mentioned, the spectrum of the \textit{Fermi} GeV excess can be
significantly affected by the uncertainties in the modeling of the Galactic
diffuse emission (which, along the line-of-sight, is typically a factor of a
few larger than the excess intensity).  In general, the relevant diffuse
foregrounds/backgrounds result from three processes: (1)~the ``$\pi^{0}$
emission", consisting of gamma rays from boosted neutral mesons (mainly
$\pi^{0}$s) that are produced when CR nucleons have inelastic collisions with
the interstellar gas, (2)~the bremsstrahlung radiation of CR electrons when
they scatter off those same interstellar gasses, and (3)~the ICS, in which CR
electrons up-scatter cosmic microwave background and interstellar radiation
field photons to gamma-ray energies. The first two contributions trace with
good accuracy the Galactic gas distribution and, as a consequence, they are
filamentary in their morphology. The ICS component, instead, is much more
diffuse and could potentially contaminate or even mimic diffuse signals from DM
annihilation.  

In most of the previous studies on the GeV excess \cite{Abazajian:2012pn,
Hooper:2013rwa, Gordon:2013vta, Abazajian:2014fta, Daylan:2014rsa} the
contribution from the Galactic diffuse emission has been modeled by using the
\PV\ model provided by the \Fermi-LAT
Collaboration.\footnote{\url{http://fermi.gsfc.nasa.
gov/ssc/data/P6V11/access/lat/ring_for_FSSC_final4.pdf}} This model was
originally developed to subtract the diffuse gamma-ray background for
point-like source emission studies, and its authors explicitly warn against
using it for the study of extended diffuse contributions.
Ref.~\cite{Calore:2014xka} showed that the \PV~model has an unusually hard ICS
component at energies above 10~GeV (this is not apparent on first sight, since
ICS, $\pi^0$ and bremsstrahlung components are not separate in this model).  In
template regression analyses, this is likely to lead to over-subtracting the
emission above 10~GeV, leading to artificial cutoffs in the GeV excess template
at these energies.

\medskip

In Fig.~\ref{fig:flux}, we show the energy spectrum of the \textit{Fermi} GeV
excess as derived in Ref.~\cite{Calore:2014xka}, including systematic and
statistical errors, compared to various DM annihilation spectra that we will
discuss further below.  In contrast to previous analyses, we find a clear
\emph{power-law like tail} at energies above 10 GeV.  However,
foreground/background model uncertainties introduce large uncertainties that
are \emph{correlated between the energy bins}.  Their effect on the fitted
spectrum is rather simple to understand and illustrated in the various panels
of Fig.~\ref{fig:flux}.  The main foreground/background components are ICS,
$\pi^0$ and bremsstrahlung.  At first order, the modeling of these components can be off in their
normalization or their slope, leading to residuals in the fit to the data that
are partially absorbed by the  \textit{Fermi} GeV excess template.
Ref.~\cite{Calore:2014xka} estimated the size of this effect from a study of
residuals along the Galactic disk, and showed that it can lead to a broadening
or narrowing of the \textit{Fermi} GeV excess spectrum, as shown in
Fig.~\ref{fig:flux}.  An immediate implication is that, in light of these
uncertainties, the \textit{Fermi} GeV excess spectrum can be fit reasonably
well with a broken power-law and different spectra from DM annihilation
models.\footnote{ Ref.~\cite{Calore:2014xka} also estimated the uncertainties
of the \emph{low-energy (sub-GeV) tail} of the spectrum.  These uncertainties
are mostly coming from the masking of point sources.  The corresponding
increase of the errors is shown in Fig.~\ref{fig:flux}.  At the lowest
energies, only upper limits on the flux can be derived.}

The impact of foreground/background model uncertainties on the \textit{Fermi}
GeV excess spectrum can be parametrized in terms of the covariance matrix of
the flux uncertainties.  The principal components of the covariance matrix
reflect the above background variations, which we found to be at the few
percent level~\cite{Calore:2014xka}.  Fits to the \textit{Fermi} GeV excess
spectrum are then performed using this simple $\chi^2$ function,
\begin{equation}
    \chi^2 = \sum_{ij} (\mu_i - f_i) \Sigma^{-1}_{ij} (\mu_j-f_j)\;,
\end{equation}
where $\mu_i$ and $f_i$ are the modeled and the measured flux in the $i^{\rm
th}$ energy bin, and $\Sigma$ is the covariance matrix.

Best-fit parameters and their uncertainties are then as usual derived by
minimizing the $\chi^2$ function w.r.t.~all model parameters, and determining
the $\Delta \chi^2$ contours while profiling over the other parameters to infer
confidence regions.

\section{Implications for models of annihilating dark matter}
\label{sec:results}

In typical DM scenarios, gamma rays are produced via a variety of mechanisms.
As DM particles annihilate, they produce Standard Model (SM) particles such as
quarks, gluons, gauge bosons and leptons. These SM particles then hadronize
and/or decay producing lighter mesons that give rise to a continuous spectrum
of gamma rays.  Moreover, $\mathcal{O}(\alpha_{\mathrm{EM}})$ corrections to
the two-body final state annihilation process generate gamma rays when an
additional photon or an SU(2) gauge boson is emitted.  Finally, at the loop
level, gamma-ray lines are expected from generic WIMP models.  In particular,
gamma rays from loop processes or emitted via electromagnetic virtual internal
bremsstrahlung (VIB) and final state radiation (FSR) give hard spectra with
evident cutoffs at the mass threshold, although suppressed in intensity.
Both, higher order correction photons and the continuous spectrum, are emitted
where the annihilation takes place and thus probe directly the annihilation
rate profile. Typically, all these components are referred to as \emph{prompt
emission}.

The gamma-ray differential intensity (with units
$\mathrm{GeV}^{-1}\,\mathrm{cm}^{-2}\,\mathrm{s}^{-1}\,\mathrm{sr}^{-1}$) from
the annihilation of self-conjugate DM $\chi$ is
\begin{equation}
    \frac{dN}{dE}=\sum_f\frac{\svf}{8\pi\, m_{\chi}^2} 
    \frac{d N^f_{\gamma}}{dE}\int_{\rm{l.o.s}}ds\, \rho^2(r(s,\psi))\;,
\end{equation}
where the sum extends over all possible annihilation channels with final state
$f$, $\svf$ is the annihilation cross-section and $dN^f_{\gamma}/dE$ is the DM
prompt gamma-ray spectrum per annihilation to final state $f$.  In this work,
the DM prompt emission spectra for all channels except $u$, $d$ and $s$ quarks
(generically $\bar{q}q$) and $hh$ are computed from the tabulated spectra
provided by \texttt{DarkSUSY}~\cite{Gondolo:2004sc}, which, in turn, derives
the $dN^f_{\gamma}/dE$ from \texttt{PYTHIA~6.4}~\cite{Sjostrand:2006za}.  We
use the $\bar{q}q$ and $hh$ spectra from \PPC~\cite{Cirelli:2010xx} (as are not
included in the \texttt{DarkSUSY} tables), which makes use of
\texttt{PYTHIA~8.135}~\cite{Sjostrand:2007gs}.  For annihilation to bosons
($W$, $Z$ and $h$) and $t$ quarks, we checked that the interpolation at mass
threshold agrees with our own results from \texttt{PYTHIA~8.186}. 

In addition to gamma rays, CR electrons and positrons are produced as final (stable) products
of DM annihilations. These CR electrons/positrons, like all other electrons/positrons propagate in
the Galaxy and produce ICS and bremsstrahlung emission.\footnote{CR $p$ and
$\bar{p}$ from DM annihilations can also give their own $\pi^{0}$ emission of
DM origin, but are suppressed from the $\bar{p}/p$ measurements already by at
least five orders of magnitude compared to the conventional Galactic diffuse
$\pi^{0}$ emission.} Generally, the ICS emission is expected to be more
important for DM models with significant branching ratios to (light) leptons.
Therefore we separate our discussion to first address the cases when ICS
emission can be safely ignored, before discussing in detail ICS emission for
annihilation to leptons.

\subsection{Single annihilation channels without ICS}
\label{sec:100BR}

\begin{figure}
    \begin{center}
        \includegraphics[width=0.9\linewidth]{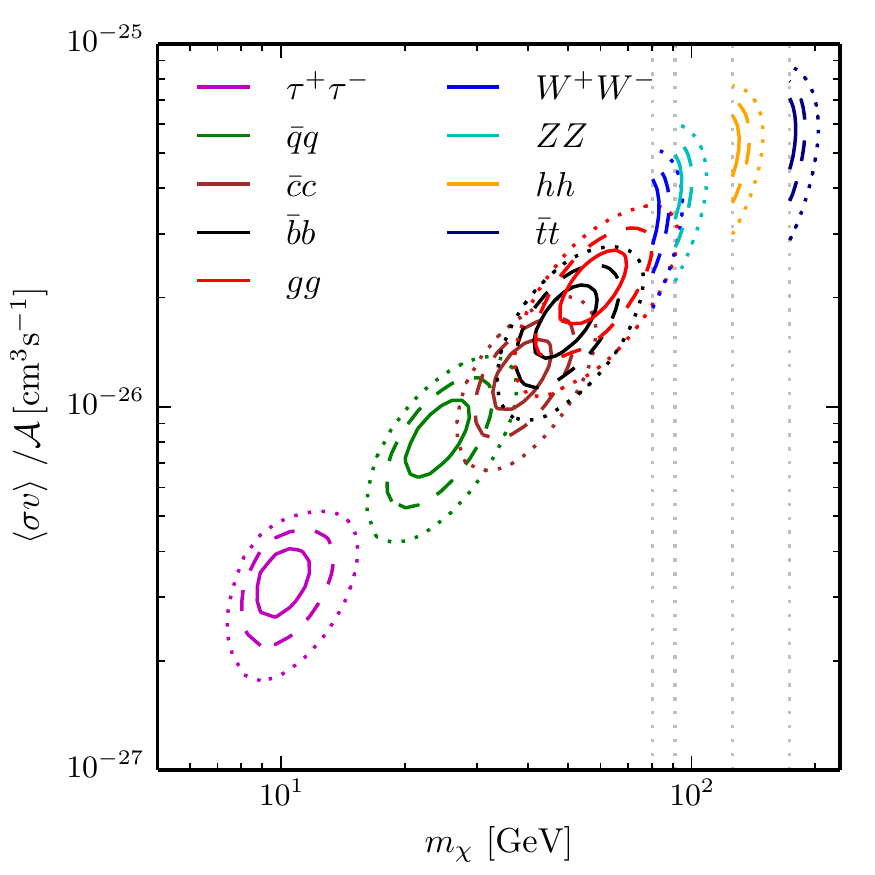}
    \end{center}
    \caption{Preferred DM mass and annihilation cross-section (1, 2 and
    3~$\sigma$ contours) for all single channel final states where ICS emission
    can be safely ignored.  \emph{Vertical gray lines} refer to the $W$, $Z$, $h$ and
    $t$ mass thresholds.  The $p$-values for annihilation to pure $W^+W^-$,
    $ZZ$ and $\bar{t}t$ final states are below 0.05, indicating that the fit is
    poor for these channels; see Tab.~\ref{tab:fitResults}.  Uncertainties in
    the DM halo of the Milky Way are parametrized and bracketed by $\mathcal{A} = [0.17,5.3]$,
    see Sec.~\ref{sec:dwarfs}. The results shown here refer to  $\mathcal{A} =
    1$.}
    \label{fig:100BR}
\end{figure}

\begin{table}
    \small
    \begin{center}
        \begin{tabular}{cccccc}
            \toprule
           Channel &  \pbox{5cm}{ $\langle\sigma v\rangle$ \\ (10$^{-26}${\rm\ cm$^3$\, s$^{-1}$})} & \pbox{5cm}{$m_\chi$ \\ (GeV)}  & $\chi^2_{\rm min}$ & $p$-value\\[3pt] \colrule
            $\bar{q}q$  &  $ 0.83_{-0.13}^{+0.15}$  &  $  23.8_{-2.6}^{+3.2}$ & 26.7 &   0.22 \\[3pt]
            $\bar{c}c$ &   $ 1.24_{-0.15}^{+0.15}$ &  $38.2_{-3.9}^{+4.7}$ & 23.6  & 0.37 \\[3pt]
            $\bar{b}b$ &   $1.75_{-0.26}^{+0.28} $&  $48.7_{-5.2}^{+6.4}$ & 23.9 &  0.35 \\[3pt]
            $\bar{t}t$ &   $5.8^{+0.8}_{-0.8} $&  $173.3^{+2.8}_{-0}$ & 43.9 & 0.003 \\[3pt]
            $ gg$ &   $ 2.16_{-0.32}^{+0.35}$ &  $ 57.5_{-6.3}^{+7.5}$ & 24.5  & 0.32 \\[3pt]
            $W^+W^-$ &  $ 3.52_{-0.48}^{+0.48}$ & $ 80.4^{+1.3}_{-0}$ & 36.7 & 0.026 \\[3pt]
            $ZZ$ &  $ 4.12_{-0.55}^{+0.55}$ &  $91.2^{+1.53}_{-0} $ &  35.3 & 0.036 \\[3pt]
            $hh$ &  $ 5.33_{-0.68}^{+0.68}$ & $ 125.7^{+3.1}_{-0}$ & 29.5 & 0.13\\[3pt]
            $\tau^+\tau^-$ &  $0.337_{-0.048}^{+0.047}$ &  $ 9.96_{-0.91}^{+1.05}$ & 33.5 & 0.055 \\[3pt]
           $ \big[\, \mu^+\mu^- $ &  $1.57_{-0.23}^{+0.23}$ & $ 5.23_{-0.27}^{+0.22}$ & 43.9  & $0.0036\big]_{\cancel{\text{ICS}}}$\\[3pt]
            \botrule
        \end{tabular}
    \end{center}
    \caption{Results of spectral fits to the \textit{Fermi} GeV excess emission
    as shown in Fig.~\ref{fig:flux}, together with $\pm1\sigma$ errors (which
    include statistical as well as model uncertainties, see text).  We also
    show the corresponding $p$-value.  Annihilation into $\bar{q}q$,
    $\bar{c}c$, $\bar{b}b$, $gg$ and $hh$ all give fits that are compatible
    with the observed spectrum.  There is also a narrow mass where annihilation
    into $\tau^+\tau^-$ is \emph{not} excluded with 95\%~CL significance.
    Annihilation to pure $W^+W^-$, $ZZ$ and $\bar{t}t$ is excluded at 95\%~CL,
    as is the $\mu^+ \mu^-$ spectrum \emph{without} ICS emission
    ($\cancel{\text{ICS}}$).  Bosons masses are from the PDG
    live~\cite{Agashe:2014kda}.}
    \label{tab:fitResults}
\end{table}

We first discuss annihilation to pure two-body annihilation states for the
cases when ICS emission can be safely ignored. This turns out to be all cases
except annihilation to electrons and muons.  In Fig.~\ref{fig:100BR} we show
the best-fit annihilation cross-section and DM mass for all other two-body
annihilation states involving SM fermions and bosons. The results are also
summarized in Tab.~\ref{tab:fitResults}, where we furthermore give the
$p$-value of the fit as a proxy for the goodness-of-fit.  As with previous
analyses, we find that annihilation to gluons and quark final states
$\bar{q}q$, $\bar{c}c$ and $\bar{b}b$, provides a good fit. In the case of the
canonical $\bar{b}b$ final states, we find slightly higher masses are preferred
compared to previous analyses, see \textit{e.g.}~Refs.~\cite{Gordon:2013vta,
Abazajian:2014fta,Daylan:2014rsa}.  This is because of the additional
uncertainty in the high-energy tail of the energy spectrum that is allowed for
in this analysis.  The highest mass to $\bar{b}b$ final states that still gives
a good fit (with a $p$-$\mathrm{value}>0.05$) is 73.9~GeV.

As the tail of the spectrum extends to higher energy, we also consider
annihilation to on-shell $\bar{t}t$ and SM bosons. For $\bar{t}t$, we find that
the fit is poor because the DM spectrum peaks at too high an energy
($\sim4.5$~GeV rather than the observed peak at 1--3~GeV). As the $p$-value is
very low for this channel, we do not consider it further.  Pure annihilation to
pairs of $W$ and $Z$ gauge bosons are also excluded at a little over $95\%$~CL
significance.  However, perhaps surprisingly, annihilation to pairs of on-shell
Higgs bosons (colloquially referred to as ``Higgs in
Space"~\cite{Jackson:2009kg}) produce a rather good fit, so long as $h$ is
produced close to rest.  This is analogous to the scenario studied
in Ref.~\cite{Bernal:2012cd} in a different context. One interesting feature of this
channel is the gamma-ray line at $m_{\chi}/2\simeq63$~GeV from $h$ decay to two
photons. This is clearly visible in the central panel of Fig.~\ref{fig:flux}.
The branching ratio for $h\rightarrow\gamma\gamma$ is $2.3 \times 10^{-3}$.
Following Tab.~\ref{tab:fitResults}, this implies a partial annihilation
cross-section into four photons with $m_\chi/2$ energy of $\langle\sigma
v\rangle_{\gamma\gamma\gamma\gamma} \simeq 1.2\times10^{-28}\cm^3\s^{-1}$.
Relevant limits from gamma-ray line searches can be found for example in
Ref.~\cite{Ackermann:2013uma} (see also Ref.~\cite{Weniger:2012tx}).  For a
contracted NFW profile (rescaled to $\gamma=1.26$), the limit for 125.7/2~GeV
mass DM particles annihilating into two photons with energy $125.7/2$~GeV is
$\langle\sigma v\rangle_{\gamma\gamma}\lesssim4.2\times10^{-29}\rm\cm^3\s^{-1}$
(at 95\% CL).  The relevant limit in our case is that $\langle\sigma
v\rangle_{\chi\chi\to hh\to\gamma\gamma\gamma\gamma}\lesssim
8.4\times10^{-29}\rm\cm^3\s^{-1}$: there is a factor $2$ because there are four
$\gamma$ in each annihilation instead of two, but this is compensated by a
factor $1/4$ from the reduction in the DM number density because, to produce
photons with the same energy, the DM must be twice as heavy in
$\chi\chi\rightarrow \gamma\gamma\gamma\gamma$ compared to $\chi\chi\rightarrow
\gamma\gamma$.  We find that $\langle\sigma
v\rangle_{\gamma\gamma\gamma\gamma}$ is therefore just below current limits.
It should be remembered that if the Higgs particles are not produced exactly at
rest, the lines are somewhat broadened, which reduces the sensitivity of line
searches~\cite{Ibarra:2012dw}.

We next turn to consider annihilation to leptons. Owing to the larger
foreground uncertainties in this analysis, we find that there is a small mass
window where $\tau^+\tau^-$ final state has a $p$--value larger than 0.05 (from
about 9.4 GeV up to 10.5 GeV).

For completeness, we also list in Tab.~\ref{tab:fitResults} the result of our
spectral fit to $\mu^+\mu^-$ final states \emph{without accounting for ICS
emission}.

Finally, we remind the reader that the quoted cross-sections assume the
Milky Way halo parameters detailed in Sec.~\ref{sec:bkg}.
These halo parameters are not well known and as we will 
discuss below in Sec.~\ref{sec:dwarfs}, dynamical and
microlensing constraints on the halo parameters (from~\cite{Iocco:2011jz})
translate to about a factor five uncertainty in the cross-section in both
directions.

\subsection{Single annihilation channels with ICS}
\label{sec:ICS}

\begin{figure}
    \begin{center}
        \includegraphics[width=0.9\linewidth]{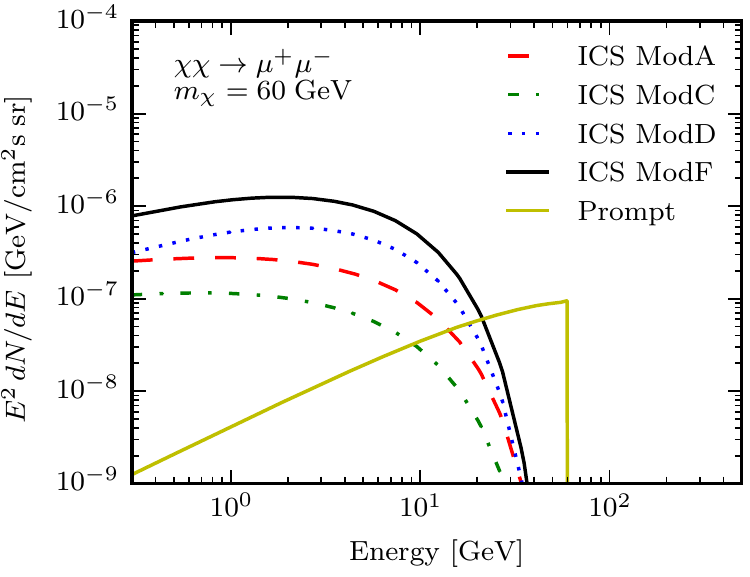}
    \end{center}
    \caption{The ICS emission spectrum from propagation models A, C, D and F
    (see Ref.~\cite{Calore:2014xka}) for a DM particle of 60 GeV annihilating
    to $\mu^{+}\mu^-$ with thermal cross-section. Fluxes are averaged over a
    $40^\circ\times40^\circ$ ROI centered on the GC, with $|b|<2^\circ$ masked.
    For comparison, we also show the prompt component of that channel, which is
    dominated by final state radiation.}
    \label{fig:ICSassump}
\end{figure}

\begin{table}
    \small
    \begin{center}
        \begin{tabular}{cccccc}
            \toprule
           Diffuse Model &  \pbox{5cm}{ $\langle\sigma v\rangle$ \\ (10$^{-26}${\rm\ cm$^3$\, s$^{-1}$})} & \pbox{5cm}{$m_\chi$ \\ (GeV)}  & $\chi^2_{\rm min}$ & $p$-value\\[3pt] \colrule
            A  &   $  12.4_{-1.6}^{+1.6}$ & $ 71.2_{-4.8}^{+5.6}$ &34.4 &  0.04 \\[3pt]
            C &   $11.8_{-3.3}^{+3.3}$ & $ 75.2_{-8.1}^{+7.9}$ & 77.5  & $\ll 10^{-3}$ \\[3pt] 
            D &   $3.56_{-0.44}^{+0.44}$ & $ 57.4_{-4.1}^{+4.6}$ & 23.9 & 0.35 \\[3pt]
            F &    $1.70_{-0.22}^{+0.22}$ &$ 60.8_{-3.9}^{+5.8}$ & 28.2 & 0.17 \\[3pt]
            \botrule
        \end{tabular}
    \end{center}
    \caption{Results of spectral fits to the \textit{Fermi} GeV excess emission
    for 100\% annihilation into $\mu^+ \mu^-$, with ICS emission modeled
    according to Galactic diffuse models A, C, D and F (see
    Ref.~\cite{Calore:2014xka}).  The $\pm1\sigma$ errors include statistical as
    well as model uncertainties, see text.  We also show the minimum $\chi^2$,
    and the corresponding $p$-value.}
    \label{tab:fitResultsICS}
\end{table}

\begin{figure*}
    \begin{center}
        \includegraphics[width=0.8\linewidth]{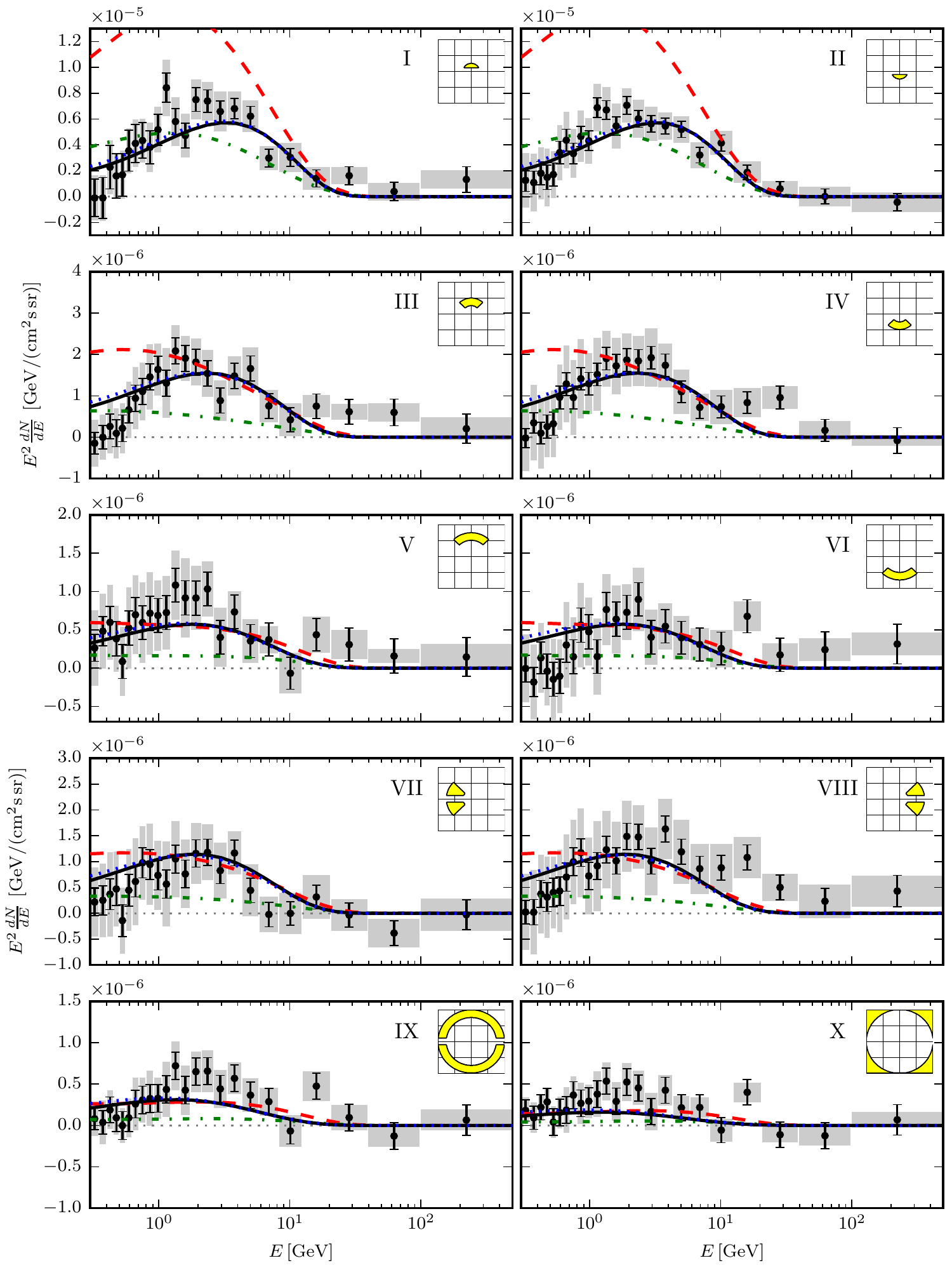}
    \end{center}
    \caption{For the same mass and cross-section as in
    Tab.~\ref{tab:fitResultsICS} the DM signal versus the gamma-ray
    \textit{Fermi} GeV excess data for the ten sub-regions of
    Ref.~\cite{Calore:2014xka} and for the four diffuse emission models adopted
    (same color/line style as in Fig.~\ref{fig:ICSassump}). In the case of
    model A (\emph{red dashed line}), while averaged over the entire ROI, the
    gamma-ray DM signal from the specific choice of mass and cross-section for
    this channel provides a good fit, once further scrutinized to the ten
    sub-regions, this DM model is excluded. On the other hand, model D
    (\emph{blue dotted line}) and F (\emph{black solid line}) still provide a
    signal compatible with the measured one in each of the 10 sub-regions.  The
    insets show the geometry of the regions in a $40^\circ\times40^\circ$ box
    centered on the GC; see also Tab.~\ref{tab:ROIs}.}
    \label{fig:MuonsRegions}
\end{figure*}

ICS emission is expected to be important for DM models with significant
branching ratios to (light) leptons (see for instance
Ref.~\cite{Lacroix:2014eea} for a discussion in the context of the GeV excess
at the GC).  Yet, any DM model that has a large
branching ratio to monochromatic $e^{+}e^-$ is severely constrained by the
positron fraction data from the AMS experiment~\cite{Bergstrom:2013jra}.
Moreover, for any DM mass the annihilation channel to monochromatic $e^{+}e^-$
would lead to an ICS gamma-ray spectrum with a hard cutoff at the mass
threshold. This though is in tension with the fact that the \textit{Fermi} GeV
excess spectrum has a very broad peak at $\simeq2$~GeV, making such a model an
improbable one in the context of the \textit{Fermi} GeV excess.  Therefore, DM
models annihilating into $e^{+}e^-$ will not be studied in this work.  We
concentrate instead on the ICS signatures from DM annihilations to
$\mu^{+}\mu^-$.

For the calculation of the ICS spectrum of DM origin we use \texttt{GALPROP
v54.1.984}\footnote{\url{http://galprop.stanford.edu/}} \cite{galprop,
Galprop1}. The ICS signal depends on the assumptions with regards to the photon
targets of the interstellar radiation field and those on the energy losses and
diffusion time scales of the electrons/positrons. We use in this work four different
Galactic diffuse emission (Galactic CR propagation) models that account for the 
relevant uncertainties. These four models are
models A, C, D and F of Ref.~\cite{Calore:2014xka}.  As can be seen in
Fig.~\ref{fig:ICSassump}, these four models
give significantly different predictions (by almost an order of magnitude) for
the averaged (over our ROI) ICS DM signal.  
Finally, bremsstrahlung of DM origin is insignificant in all these
cases and thus can be ignored.\footnote{We find the ratio of 
ICS/bremsstrahlung flux to be between 10 and 100, for all the
relevant DM annihilation modes and for gamma-ray energies $<10$ GeV that
affect the spectral fits.}

\medskip
 
As for annihilation into two-body final states, the ICS emission plays an
important role for the $\mu^+\mu^-$ channel.  Annihilation into $\mu^+\mu^-$ is
excluded with high confidence level ($p$-value of 0.0036) when only the prompt
emission is considered, as can be see in Tab.~\ref{tab:fitResults}. Instead,
the inclusion of the ICS component significantly improves the fit.
Tab.~\ref{tab:fitResultsICS} quotes the best-fit parameter values and $p$-values
for annihilation into $\mu^+\mu^-$ when the ICS emission is modeled according
to the four different propagation models introduced above.  The effect of ICS
inclusion is two-fold: First, the best-fit mass range is shifted towards higher
masses ($\sim$ 60--70 GeV), while the best-fit cross-section value can vary
by about a factor of ten depending on the model.  Second, it is possible to
find models for which the goodness-of-fit is improved and can become
competitive with annihilation to $\bar{b}b$, $\bar{c}c$, light quarks
and gluons (compare models D and F for the  $\mu^+\mu^-$  channel from
Tab.~\ref{tab:fitResultsICS} to the relevant channels of
Tab.~\ref{tab:fitResults}).  The reason why such a possibility opens is that,
as shown in the lower panel of Fig.~\ref{fig:flux} and Fig.~\ref{fig:ICSassump}, the
combined ICS and prompt gamma-ray spectrum can be significantly altered with
the ICS smooth bump dominating in the fit over the FSR hard spectral feature.
This is the case for certain Galactic diffuse emission model assumptions; 
the model must allow for the injected electrons/positrons from DM annihilation to lose 
most of their energy via ICS emission. 
If CR electrons and positrons instead have large diffusion time scales (slow diffusion) and/or lose
most of their energy via the competing synchrotron losses, the ICS at higher
latitudes will be strongly suppressed.

\medskip 

For channels that have significant ICS emission, it is not enough to fit
the spectral energy distribution, but also the morphology of the signal has to
be checked.  Indeed, as mentioned above, the ICS emission is
strictly correlated with the distribution of the ambient photons. Thus, the
morphology of the emission associated with ICS emission could in principle be different
with the morphology of the GeV excess.\footnote{The prompt-only component does not
have such a drawback as a consequence of the fact that the prompt photons
directly trace the DM distribution in the Galaxy.} For this reason, in
Fig.~\ref{fig:MuonsRegions} we show the results of the fits in the ten ROIs
used in Ref.~\cite{Calore:2014xka} to characterize the morphology of the GeV
excess. The definition of the ten ROIs allowed us to study the symmetry
properties of the excess and its extension in latitude. For the sake of completeness,
we quote in Tab.~\ref{tab:ROIs} the definition of the ten ROIs as in Ref.~\cite{Calore:2014xka}.
 In each sub-region we display the GeV excess data together with the
ICS emission from $\mu^+\mu^-$ annihilation for the four Galactic diffuse
emission models A, C, D and F, and for the DM masses and cross-sections quoted in
Tab.~\ref{tab:fitResultsICS}.  Models D and F are able to reproduce the correct
morphology of the excess, while models A and C fail in this respect.
The plot is illustrative of the fact that it is possible to find propagation
models for which the morphological properties of the GeV excess are recovered.
Thus, it is not possible to exclude ICS emission from muons only on
the basis of gamma-ray data.

\begin{table}
    \small
    \begin{center}
        \begin{tabular}{cc}
            \toprule
                ROI & Definition  \\\colrule
                I, II & $\sqrt{\ell^2 + b^2} < 5^\circ$, $\pm b>2^{\circ}$  \\
                III, IV & $5^\circ<\sqrt{\ell^2 + b^2} < 10^\circ$, $\pm b>|\ell|$  \\
                V, VI & $10^\circ<\sqrt{\ell^2 + b^2} < 15^\circ$, $\pm b>|\ell|$ \\
                VII, VIII & $5^\circ<\sqrt{\ell^2 + b^2} < 15^\circ$, $\pm\ell>|b|$  \\
                IX & $15^\circ<\sqrt{\ell^2 + b^2} < 20^\circ$ \\
                X & $20^\circ<\sqrt{\ell^2 + b^2}$, $|b|<20^{\circ}$, $|\ell|<20^{\circ}$  \\
            \botrule
        \end{tabular}
    \end{center}
    \caption{Definition of the ten ROIs used in Ref.~\cite{Calore:2014xka}
            for the morphological analysis of the excess. Table adapted from 
             Ref.~\cite{Calore:2014xka}.}
    \label{tab:ROIs}
\end{table}

Other important constraints on DM annihilating to muons come from the CR
positron fraction measured by AMS-02~\cite{Bergstrom:2013jra}.  For model F the
combination of best-fit cross-section and mass in
Tab.~\ref{tab:fitResultsICS} is still allowed at $95\%$~CL, once
uncertainties on the local DM density and local CR electrons energy losses are
taken into account.  Instead, models A and C are already in strong tension.
An appealing feature of this channel is that anti-proton
 constraints~\cite{Bringmann:2014lpa, Cirelli:2014lwa, Evoli:2011id,
Cholis:2010xb, Donato:2008jk, Hooper:2014ysa} do not apply.
Before drawing any final conclusion about the possibility of having
annihilation into muons it is also important to test synchrotron
radiation. This is beyond the scope of the present paper and will be addressed in
a dedicated work.\footnote{It may also be interesting to include the
bremsstrahlung emission in some cases~\cite{BoehmFuture}.} 

We briefly mention why ICS emission is not important for annihilation into
 light quarks or $\tau^+\tau^-$.
For annihilations into light quarks and
gluons, only $\sim 1/6$ of the available energy per annihilation goes to
$e^{+}e^-$ after all the hadronization and decay processes have occurred.
Moreover the spectra of these $e^{+}e^-$ tend to be soft at injection, resulting
in an ICS signal that is subdominant compared to the gamma-ray prompt emission
signal around the energies of the \textit{Fermi} GeV excess.  
For direct annihilation to $\tau^+\tau^-$, while a significant portion of
the annihilation power does go into $e^{+}e^-$, the prompt gamma-ray emission 
has a very prominent spectral bump, that cannot
be ``smoothed out" significantly by including the ICS contribution. 
In these cases, we have checked
that including the ICS emission impacts the
best-fit mass and cross-section in Tab.~\ref{tab:fitResults} and their respective 1, 2 
or 3~$\sigma$ ranges in Fig.~\ref{fig:100BR} by no more than~$5\%$.

Finally we notice that for a point-like source of high energy
electrons located at the GC, either at a steady rate or for a sequence of burst-like events with
a time-scale separation between the events of $\sim \mathcal{O}(100)$ kyr or
less, the simultaneous explanation of the spectrum in the entire ROI of
Ref.~\cite{Calore:2014xka} ($| l | <  20^{\circ}$, $2^{\circ} < |b|
<20^{\circ}$) and its ten sub-regions is going to be challenging. We expect
for the ICS signature, even if it fits the entire ROI, it will overshoot the data in
the inner sub-regions (mainly regions I and II) and undershoot the outer
sub-regions data, much like in Galactic diffuse emission model A of Fig.~\ref{fig:MuonsRegions}. 

\subsection{Mixed annihilation channels}

\begin{figure*}
    \begin{center}
        \includegraphics[width=0.33\linewidth]{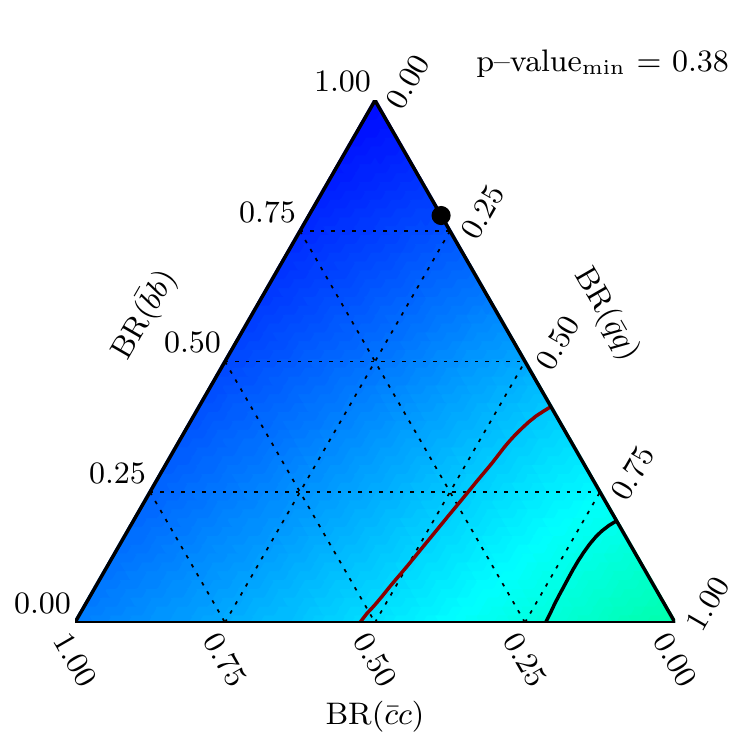} 
        \includegraphics[width=0.33\linewidth]{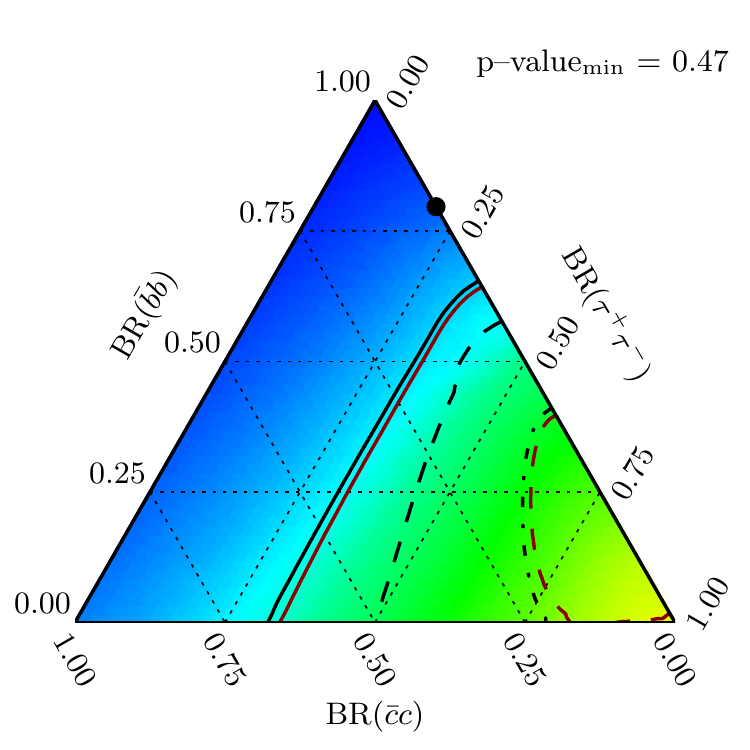} 
        \includegraphics[width=0.33\linewidth]{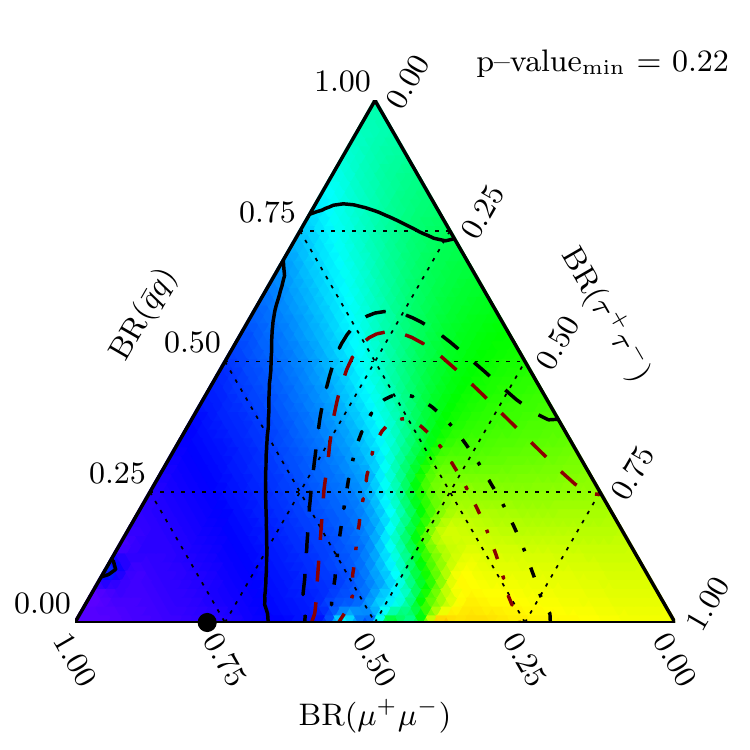} 
        \includegraphics[width=0.33\linewidth]{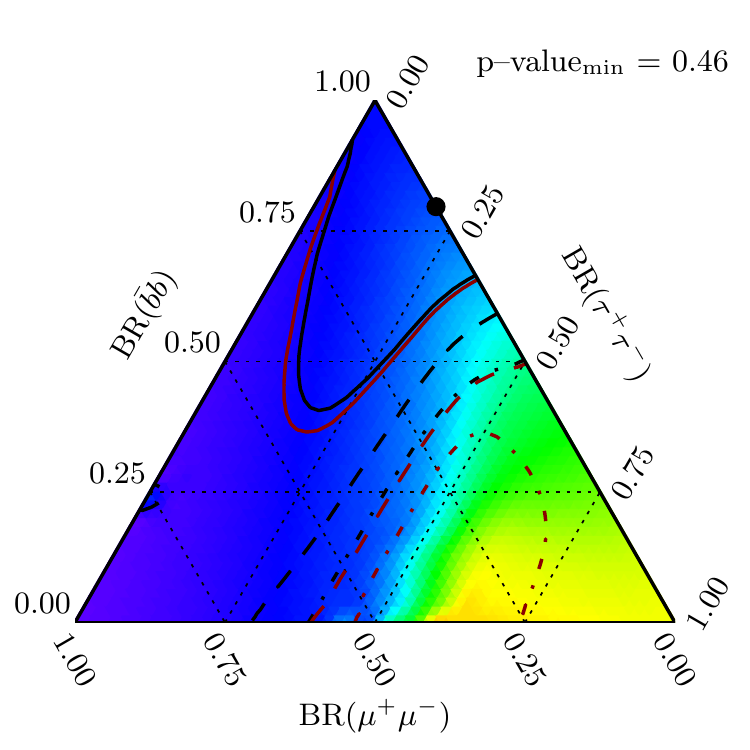} 
        \includegraphics[width=0.35\linewidth]{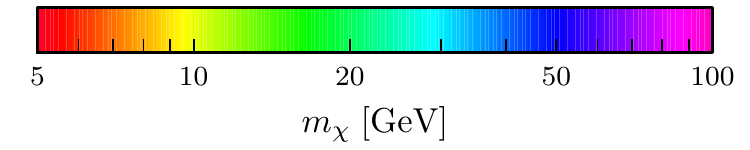}
    \end{center}
    \caption{Constraints on the branching ratio (BR) to mixed final states that
    include quarks and leptons. We marginalize over the DM mass and the total
    annihilation cross-section. The angles of each triangle represent
    annihilations to a pure channel, with the mass and cross-section being the
    best fit values given in Tabs.~\ref{tab:fitResults}
    and~\ref{tab:fitResultsICS} (model F).  The \emph{black dot} in each
    plot corresponds to the best-fit point (we give the $p$-value here), the
    \emph{solid}, \emph{dashed} and \emph{dot-dashed black lines} show the $1$,
    $2$, and $3~\sigma$ contours about the best-fit point, and the
    \emph{solid}, \emph{dashed} and \emph{dot-dashed red lines} indicate
    $p$-value contours of $0.32$, $0.05$ and $0.01$ respectively. Any
    combination of light quarks always results in a good fit. 
    The BR to $\tau^+ \tau^-$ can be substantial, with values over
    $\sim50\%$ allowed at 2$\sigma$. 
    Owing to the inclusion of ICS emission, any value of
    $\mathrm{BR}(\mu^+ \mu^-)$ results in a good fit (\cf bottom panels) when
     some fraction of $\bar{q}q$, $\bar{b}b$ or $\tau^+ \tau^-$ is also
    included.  In each panel the background coloring refers to the best-mass
    range as indicated by the color bar. Masses in the range 35--60 GeV lie
    inside the best-fit regions for all the shown combinations. }
    \label{fig:hadronicBR}
\end{figure*}

\begin{figure}
    \begin{center}
       \includegraphics[width=0.64\linewidth]{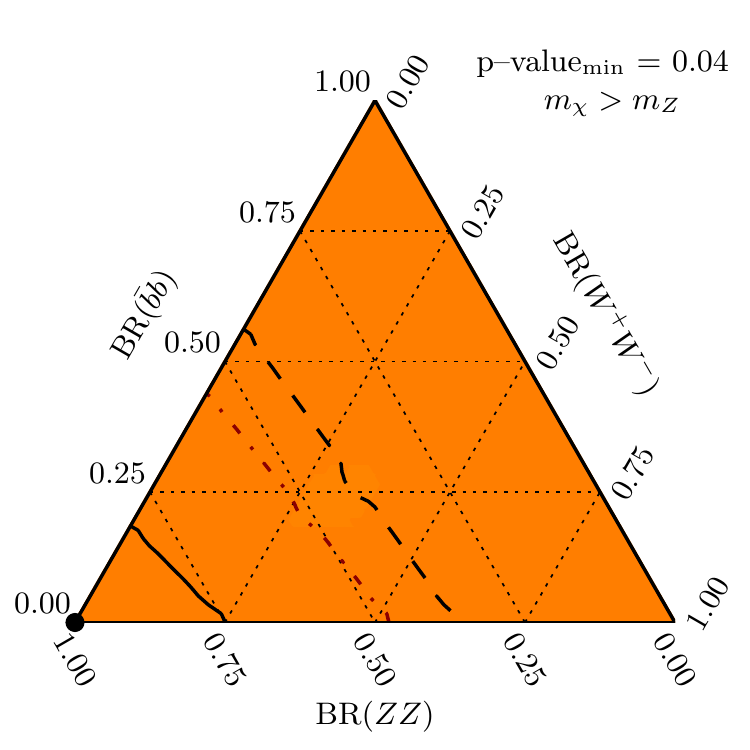}
       \includegraphics[width=0.64\linewidth]{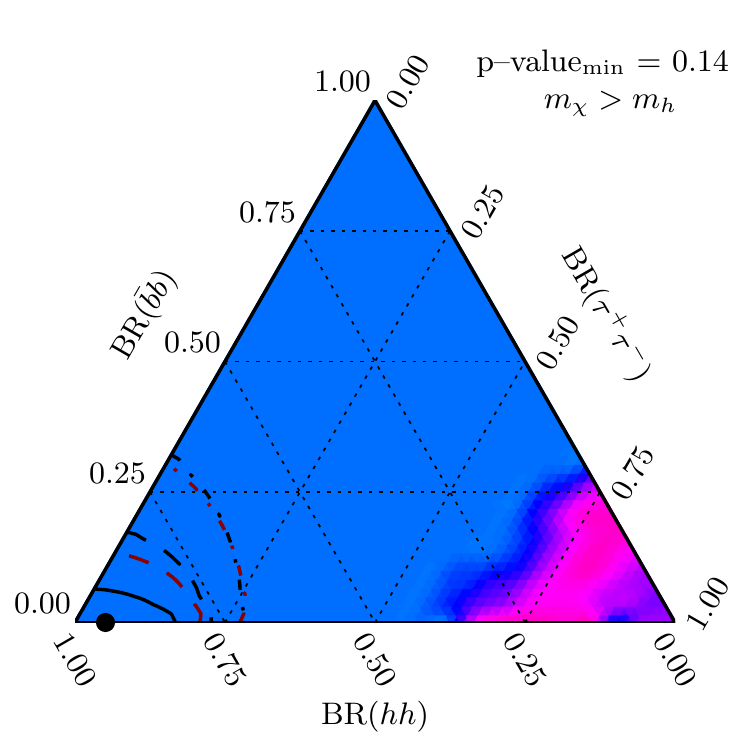}
       \includegraphics[width=0.64\linewidth]{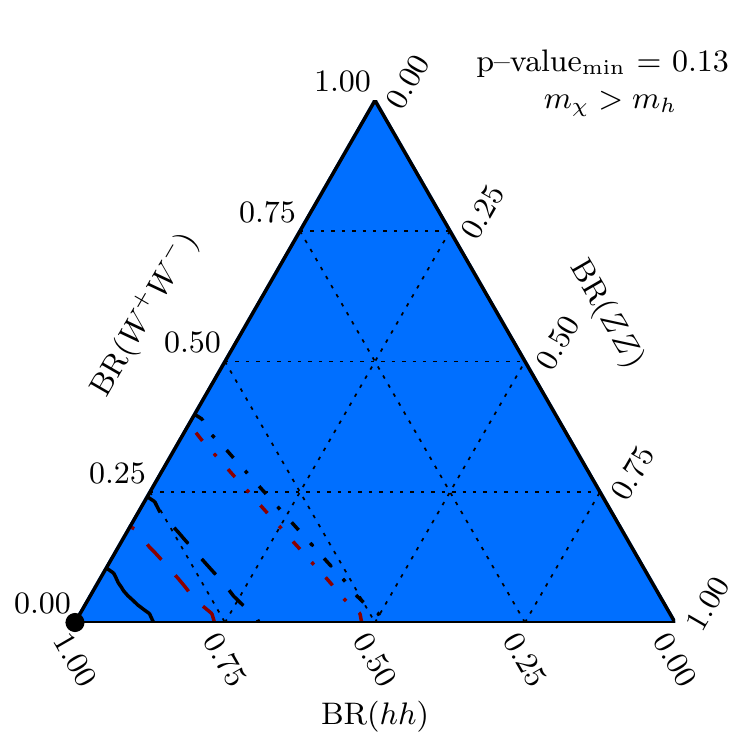}    
       \includegraphics[width=0.7\linewidth]{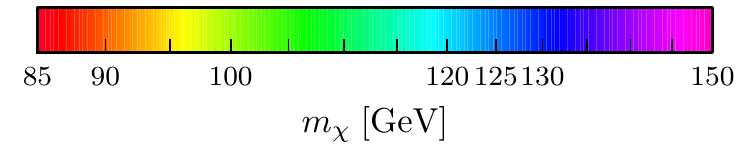}   
    \end{center}
    \caption{Same as Fig.~\ref{fig:hadronicBR}, but for a selection of mixed
    final states that include SU(2) gauge bosons and the Higgs channel on
    shell.  \emph{Upper panel:} We show the combination of heavy gauge bosons,
    $W$ and $Z$, together with the possibility to annihilate into $b$ quark final
    states. Given the imposed constraint $m_{\chi} > m_{Z}$, the best-fit
    region is forced to be close to the pure annihilation to $ZZ$.
    \emph{Central panel:} We show a combination of $b$ quarks, $\tau$-leptons and
    $hh$.  As above the constraint from the Higgs channel on the DM mass,
    $m_{\chi} > m_{h}$, causes the best-fit region to be close to the pure
    annihilation to $hh$. We notice that, when a substantial fraction of
    $\tau^+\tau^-$ is allowed, together with the mass constraint, the fit is
    very bad.  \emph{Lower panel:} We show the combination of $W^{+}W^{-}$,
    $ZZ$ and $hh$ annihilation states.}
    \label{fig:mixedh}
\end{figure}

The discussion so far has focused on annihilation into a single channel of
final states.  In a realistic model, DM will likely annihilate into a variety
of channels with different branching ratios, defined as
$\mathrm{BR}(\bar{f}f)=\svf/\sum_f \svf$ where the sum extends over all
available channels. For instance, all but two of the models in
Refs.~\cite{Logan:2010nw, Buckley:2010ve, Zhu:2011dz, Marshall:2011mm,
Boucenna:2011hy, Buckley:2011mm, Anchordoqui:2013pta, Buckley:2013sca,
Hagiwara:2013qya, Okada:2013bna, Huang:2013apa, Modak:2013jya, Boehm:2014hva,
Alves:2014yha,Berlin:2014tja,Agrawal:2014una,Izaguirre:2014vva,
Cerdeno:2014cda, Ipek:2014gua,Boehm:2014bia,Ko:2014gha, Abdullah:2014lla,
Ghosh:2014pwa, Martin:2014sxa, Basak:2014sza, Berlin:2014pya, Cline:2014dwa,
Han:2014nba, Detmold:2014qqa, Wang:2014elb, Chang:2014lxa, Arina:2014yna,
Cheung:2014lqa, McDermott:2014rqa, Huang:2014cla,
Balazs:2014jla,Ko:2014loa,Okada:2014usa,Ghorbani:2014qpa,
Banik:2014eda,Borah:2014ska,Cahill-Rowley:2014ora,Guo:2014gra,Freytsis:2014sua}
annihilate into multiple final states (the exception being flavored DM
models~\cite{Agrawal:2014una,Izaguirre:2014vva}, where annihilation into only
$\bar{b}b$ is possible).

Previous multi-channel fits to the GeV excess have generally focused on the
cases where $\svf \propto \{m_f^2, e_f^2, 1\!\!1\}$, where $m_f$ is the final
state mass, $e_f$ the final state electric charge and $1\!\!1$ denotes
universal couplings. These scenarios can be motivated by considering models
where the particle mediating the annihilation mixes with the SM Higgs (in
variations of two-Higgs doublet (2HDM) or Higgs portal
models~\cite{Patt:2006fw}) or from Minimal Flavor
Violation~\cite{D'Ambrosio:2002ex} (in the case $\svf \propto m_f^2$), where a
vector mediator kinetically mixes with electromagnetism (when $\svf \propto e_f^2$)
or where the couplings are assumed universal as a simplifying assumption (when
$\svf \propto1\!\!1$).  Here however, we remain more agnostic to the allowed
final states. We do this for two reasons: Firstly, models often predict
deviations from the exact relations $\svf \propto \{m_f^2, e_f^2, 1\!\!1\}$.
Secondly, not all models have been explored so we do not want to over restrict
ourselves.

\medskip 

We therefore show in Figs.~\ref{fig:hadronicBR} and~\ref{fig:mixedh} triangle
plots with fits to three final state channels. The plots are such that the
branching ratios (BR) sum to one (as required) and we have marginalized over the DM
mass and the total annihilation cross-section. Owing to the large uncertainty
on the total cross-section from the Milky Way halo parameters (about a factor
five as we anticipated in Fig.~\ref{fig:100BR} and discussed in Sec.~\ref{sec:dwarfs}),
 we choose  to show the
DM mass that minimizes the $\chi^2$ at each point by means of the background 
coloring.

\smallskip

Fig.~\ref{fig:hadronicBR} focuses on the case of annihilation to quark and
lepton channels.  The top-left triangle is for quark-only final states (we
don't consider $gg$ as it is loop suppressed so its branching ratio is
naturally smaller). As each channel individually gives a good fit, it is no
surprise that any combination of $\bar{q}q$, $\bar{c}c$ and $\bar{b}b$ also
gives a good fit with DM in the mass range between 25 and 60~GeV.  The
top-right triangle is for heavy quark and $\tau$ final states, as would be
expected for 2HDM and Higgs-portal models with $m_{\chi}<m_t$. The best-fit
point lies close to the ratios predicted in these models
$\bar{b}b:\bar{c}c:\tau^+ \tau^-=0.87:0.08:0.05$. We also find that
$\mathrm{BR}(\tau^+\tau^-)$ can be substantial (up to around 75\%) while still
providing a good spectral fit ($p$-value $> 0.05$).  The bottom two
triangles consider annihilation channels involving muons, where we include ICS
emission assuming propagation model F.  Considering first the case of annihilation into
$\mu^+\mu^-$ and $\tau^+ \tau^-$ only (along the axis $\mathrm{BR}(\bar{q}q)=0$), we
see that a good fit can always be obtained when
$\mathrm{BR}(\mu^+\mu^-)\gtrsim0.6$ and $m_{\chi}\sim50$~GeV.  Constructing
models giving only $\mu^+\mu^-$ and $\tau^+ \tau^-$ final states may be
challenging but see Ref.~\cite{Buckley:2013sca} for a prototype. When allowing
for additional annihilation into $\bar{q}q$ or $\bar{b}b$, we find that all
values of $\mathrm{BR}(\mu^+\mu^-)$ result in a good fit.

\smallskip
 
In Fig.~\ref{fig:mixedh} we consider final states which contain at least one
heavy boson.  The single channel annihilation to $W^{+}W^-$ or $ZZ$ is just
excluded at 95\%~CL, therefore in the upper triangle we investigate whether a
combination of $\bar{b}b$ in addition to $W^{+}W^-$ and $ZZ$ improves the fit.
Unfortunately we find that this is not the case: the best-fit point remains
pure annihilation to $ZZ$. We find the same conclusion for annihilation to
other light quark anti-quark pairs (not shown).  This is expected because of
the constraints imposed on the DM mass, $m_{\chi} > m_{Z}$ (due to the
requirement of producing the heavy bosons on shell).
 
In Sec.~\ref{sec:100BR}, we found that single channel annihilation to $hh$
gives a fit that it compatible with the observed energy spectrum. In the middle
and lower triangles we investigate if the inclusion of additional quark and
lepton final states, and additional $W^{+}W^-$ and $ZZ$ final states improves
the pure $hh$ fit. A small fraction with $\bar{b}b$ provides a marginally
better fit but generally -- when adding both quark and leptons or heavy gauge
bosons -- we find that a good fit is obtained only when
$\mathrm{BR}(hh)\gtrsim0.8$ and when the DM mass is close to $m_h$ (recall that
also in this case the production of on-shell Higgs imposes $m_{\chi} > m_{h}$).
This implies that simple models such as singlet scalar
DM~\cite{Silveira:1985rk,McDonald:1993ex,Burgess:2000yq,Cline:2013gha}, which
predicts sizable branching ratios to $W^{+}W^-$, $ZZ$ and $hh$, are excluded
since there $\mathrm{BR}(W^{+}W^-)$ is largest. Building realistic models that
annihilate dominantly to $hh$ may prove challenging, but opens to new,
unexplored, possibilities.

\subsection{Annihilation to hidden sector mediators}

Up to this point we have only considered scenarios where the DM particles annihilate directly to
SM particles.  However it is also plausible that the DM first annihilates to
intermediate hidden sector mediators $\phi$ that subsequently decay to SM
particles. The mediator $\phi$ can mix with the SM Higgs or with
hypercharge/electromagnetism, allowing for a variety of possible SM states from
their decays.

These ``cascade'' annihilations produce boosted SM final states, which, depending
on the $\phi$ mass, allow for heavier DM particles than in the more
conventional scenarios discussed previously.  The case in which a general
mediator $\phi$ decays primarily to $b$ quarks has already been discussed
extensively in the literature~\cite{Boehm:2014bia,Ko:2014gha,Abdullah:2014lla,
Martin:2014sxa,Berlin:2014pya,Cline:2014dwa,Ko:2014loa}. In fact the single
channel annihilation to $hh$ can be considered in this class since, after the
$h$ is produced, it decays dominantly to $\bar{b}b$ with each $b$ having energy
$m_h/2$.  This is why a DM interpretation for this channel results in a
good-fit even though the DM mass is over twice as heavy compared with the
values for other channels.
 
Here we consider eXciting Dark Matter models (XDM)\cite{Finkbeiner:2007kk, ArkaniHamed:2008qn}.
For an earlier discussion of XDM models in the
context of the \textit{Fermi} GeV excess see~\cite{Hooper:2012cw}.  If the gauge bosons $\phi$ are
lighter than 2~GeV, the kinematically allowed final states are
$e^{+}e^-$, $\mu^{+}\mu^-$ and $\pi^{+}\pi^-$ or $\pi^{0}$s, while no anti-protons are produced, 
thus evading the current constraints \cite{Cholis:2008qq}. Such channels will produce, after all the subsequent 
cascades, boosted electrons and positrons and a subdominant contribution to FSR \cite{Cholis:2008wq}.

The $\pi^{0}$ channel can be evaded if the $\phi$ mixes with electromagnetism, thus coupling
to charge \cite{ArkaniHamed:2008qn}.  We will therefore concentrate here on the annihilation
channel $\chi \chi \to \phi \phi$, with subsequent $\phi$ decays as $\phi \to
e^{+}e^{-}$, $\phi \to \mu^{+}\mu^{-}$ or $\phi  \to \pi^{+}\pi^{-}$.\footnote{For a case where the $\pi^{0}$ modes dominate, see \cite{Liu:2014cma}.}
 
\begin{table}
    \small
    \begin{center}
        \begin{tabular}{cccccc}
            \toprule
           Channel &  \pbox{5cm}{ $\langle\sigma v\rangle$ \\ (10$^{-26}${\rm\ cm$^3$\, s$^{-1}$})} & \pbox{5cm}{$m_\chi$ \\ (GeV)}  & $\chi^2_{\rm min}$ & $p$-value\\[3pt] \colrule
            $\phi \to e^{+}e^{-}$  &  $ 0.384_{-0.051}^{+0.052}$  &  $  45.7_{-3.3}^{+3.4}$ & 31.35 &   0.09 \\[3pt]
             $\phi \to \mu^{+}\mu^{-}$ &   $ 2.90_{-0.42}^{+0.43}$ &  $91.7_{-7.5}^{+8.9}$ & 33.6  & 0.05 \\[3pt]
             $\phi \to \pi^{+}\pi^{-}$ &   $ 5.11_{-0.71}^{+0.72} $&  $124.5_{-9.8}^{+11.3}$ & 33.3 &  0.06 \\[3pt]
            \botrule
        \end{tabular}
    \end{center}
    \caption{As in Tab.~\ref{tab:fitResults}, results of spectral fits to the \textit{Fermi} 
    GeV excess emission, for DM models annihilating into light bosons $\phi$.
    The corresponding $p$-value is $ \geq0.05$ in all cases.
    A slightly better fit is provided by  $\phi \to e^{+}e^{-}$. For the ICS emission we considered the diffuse emission model F.}
    \label{tab:fitResultsXDM}
\end{table}

\begin{figure}
    \begin{center}
       \includegraphics[width=0.7\linewidth]{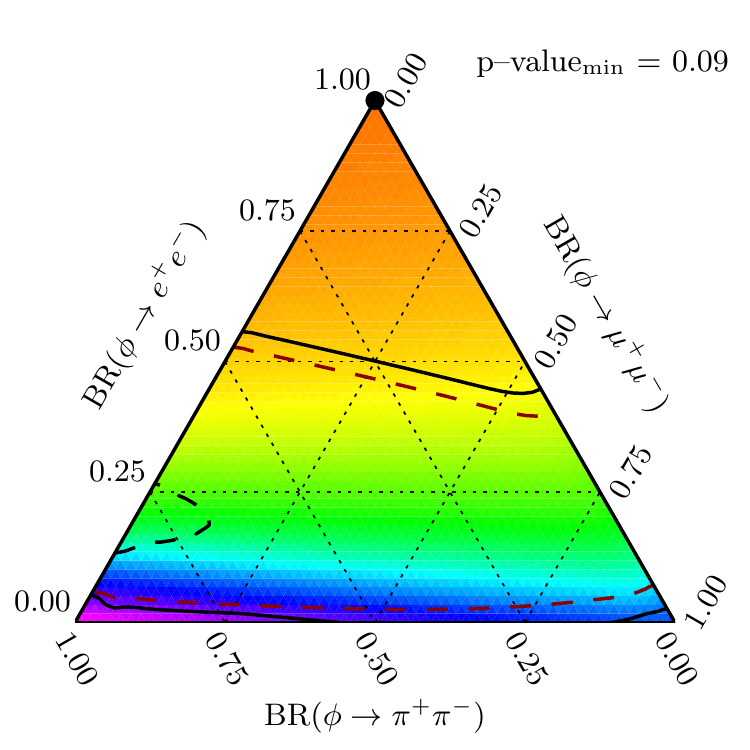}
       \includegraphics[width=0.7\linewidth]{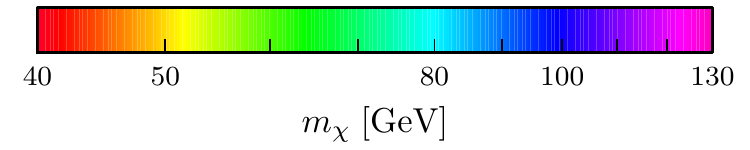}   
    \end{center}
    \caption{Same as Fig.~\ref{fig:hadronicBR}, but for annihilation into light bosons $\phi$,
    which subsequently decay to $\phi \to e^{+}e^{-}$, $\phi \to \mu^{+}\mu^{-}$
    and $\phi \to \pi^{+}\pi^{-}$. While the best fit case is for the pure case  to 
    $\phi \to e^{+}e^{-}$, at the 2$\sigma$
    level a wide variety of possible BRs and a range of masses between 45 GeV and 125 GeV 
    is allowed.}
    \label{fig:mixedXDM}
\end{figure}

As the final states contain light leptons, it is again crucial to include ICS
emission.  We do this as before using the Galactic diffuse emission model~F. 
We show the results from our spectral fits in Tab.~\ref{tab:fitResultsXDM} for single channel decay 
to each of the three possible $\phi$ decay modes: 
$e^{+}e^{-}$, $\mu^{+}\mu^{-}$ or $\pi^{+}\pi^{-}$.
We find that the best-fit case, $\phi \to e^{+}e^{-}$, suggests a mass and a cross-section
that is still allowed from AMS positron fraction limits, within their uncertainties, similarly to the case of direct DM annihilation to $\mu^{+} \mu^{-}$ discussed in Sec.~\ref{sec:ICS}.
Fig.~\ref{fig:mixedXDM} shows
the resulting triangle plot for floating BRs between the three $\phi$ decay modes,
after marginalizing over the DM mass and the
annihilation cross-section to produce $\phi \phi$. 
Again the AMS 
positron fraction limits constrain (but not severely) these possibilities. 
For reference in these calculations we have chosen $\phi$ to be a vector with a mass of
$\simeq 0.6$ GeV. Our spectral fit results do not depend on the exact value of the $\phi$
mass, as long as it remains within 0.3--1 GeV, and on whether $\phi$ is a vector or a scalar,
given the similarity of the injected electron/positron spectra into the interstellar medium from these options. 
Yet, on the 
model building side these can be important assumptions~\cite{ArkaniHamed:2008qn, Finkbeiner:2010sm}.  

\section{Current and future constraints from Dwarf Spheroidals}
\label{sec:dwarfs}

\begin{figure}
    \begin{center}
        \includegraphics[width=0.9\linewidth]{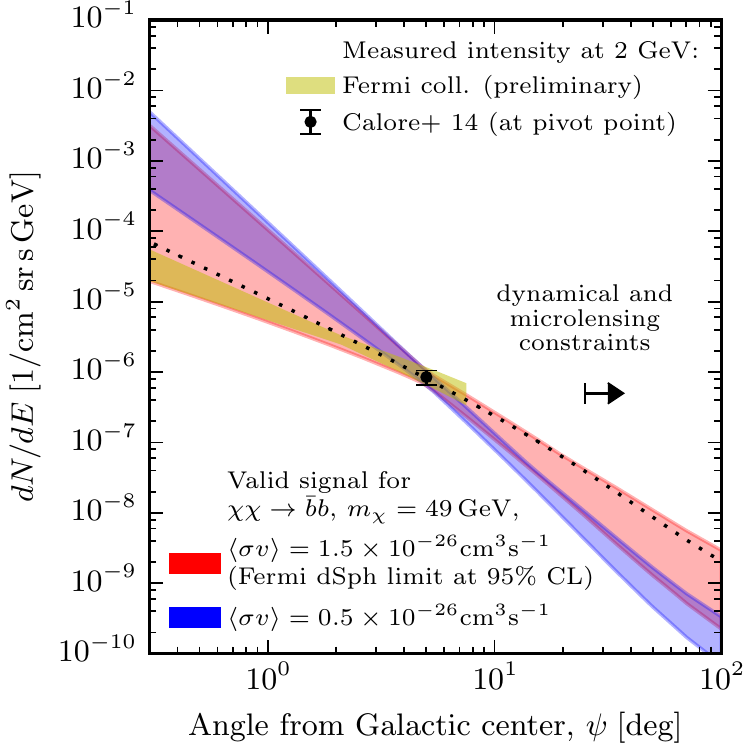}
    \end{center}
    \caption{Radial intensity profile of the \textit{Fermi} GeV excess, at 2
    GeV, \cf Fig.~\ref{fig:measured_profile}.  The \emph{black data point} refers to
    measurements from Refs.~\cite{Calore:2014xka}, the \emph{yellow band} to
    preliminary results from the \Fermi-LAT team~\cite{simonaTalk}. The \emph{dotted
    line} shows the expectations for a contracted NFW profile from
    Fig.~\ref{fig:measured_profile}.  The \emph{red} and \emph{blue bands} show -- for a
    given DM annihilation scenario -- possible signal morphologies that are
    compatible with both the measurements at $\psi=5^\circ$ as well as
    dynamical and microlensing observations from Ref.~\cite{Iocco:2011jz} (we
    concentrate on arbitrary generalized NFW profiles).  For annihilation
    cross-sections close to the current dwarf limits, the intensities
    determined by different groups (as indicated by the dotted line), lie still
    in the allowed range.}
    \label{fig:profile}
\end{figure}

The arguably most promising channel for a confirmation of the DM interpretation
of the \textit{Fermi}  GeV excess are searches for corresponding signals in
dwarf spheroidal galaxies of the Milky Way.  These observations probe
already -- for typical assumptions on the Milky Way DM halo -- DM scenarios
that could explain the \textit{Fermi}  GeV
excess~\cite{Geringer-Sameth:2014qqa, Cholis:2012am, GeringerSameth:2011iw,
Abdo:2010ex}.  The currently strongest (though still preliminary) limit on the
annihilation cross-section was presented in Ref.~\cite{brandonTalk}.  For
annihilation into $\bar bb$ final states and a DM mass of $49\GeV$, they read
$\langle \sigma v \rangle < 1.5\times10^{-26}\cm^3\s^{-1}$ at 95\% CL, which is
at face-value in mild tension with the values of the cross-section that we
derived above (see Sec.~\ref{sec:results}).  However, the link between the
GC and the dwarf signals is subject to uncertainties in the DM distribution in
the Milky Way and the DM distribution in dSphs (note that the latter
have been marginalized over in the analysis of Ref.~\cite{brandonTalk}).
Concerning the Milky Way halo, a decrease of the scale radius
$r_s$, an increased slope $\gamma$ of the inner part of the profile, or an
increased local density $\rho_\odot$ enhance the expected GC signal relative
to the signal in dwarf spheroidals. Also more cored profiles for dSphs can
reduce further their constraining power.  It is important to investigate to
what extent uncertainties in these parameters can mitigate potential tensions
between GC and dSph observations~\cite{woodTalk}.

\medskip

In Fig.~\ref{fig:profile} we show the \emph{expected} signal flux for DM
annihilation into $\bar{b}b$ final states and with $m_\chi=48.7\GeV$.  As DM
profile we adopt here the reference generalized NFW profile as above and the
cross-section is set to $\sv = 1.75\times10^{-26}\cm^3\s^{-1}$. This leads to
a signal intensity that is consistent with the results found in
Ref.~\cite{Calore:2014xka} at $\psi=5^\circ$. Note that $\psi=5^\circ$ was
found to be a good pivot point for the intensity measurement in
Ref.~\cite{Calore:2014xka}, as the flux there is relatively independent of the
adopted profile slope.  We also show the preliminary GC results by the \Fermi-LAT
Collaboration, \cf Fig.~\ref{fig:measured_profile}.

To explore the validity of measured signal profile, we generate a large set of
Milky Way DM halo models that are compatible with the microlensing and
dynamical constraints from Ref.~\cite{Iocco:2011jz} at 95\%~CL.  This set
includes DM halo models that follow a generalized NFW profile with scale radii
in the range $r_s = 10$ to $30\kpc$, and arbitrary normalization $\rho_s$ and
inner slope $\gamma$ (note that for illustrative purposes we allow also values
of $\gamma$ that would be incompatible with the \textit{Fermi}  GeV excess
measurements at the GC).  To this end, we adopt the following
method:  We derive the envelope of all density profiles that are compatible
with the left panel of Fig.~5 of Ref.~\cite{Iocco:2011jz} (which shows results
for $r_s=20\kpc$ only) in the radial range $r=2.5$ to $10\kpc$.  A model with
scale radius $r_s\neq20\kpc$ is considered to be compatible with the
observations when its profile lies within the derived envelope.

\medskip

From the set of all observationally allowed halo models we select those that
lead to a signal intensity that is consistent with the measurements at
$\psi=5^\circ$, assuming a reference cross-section $\langle \sigma
v\rangle=1.5\times10^{-26}\cm^3\s^{-1}$ (the current dSph limit at 95\%~CL) and
the above annihilation channel and DM mass.  The envelope of the corresponding
allowed signal profiles is shown by the red band in Fig.~\ref{fig:profile}.
The band contains both the signal morphology as derived for the reference
generalized NFW profile with $\gamma=1.26$, as well as with the preliminary GC
results by the \Fermi-LAT Collaboration.  \emph{We hence find that current dSph
limits on the annihilation cross-section are well consistent with a DM
interpretation of the \textit{Fermi}  GeV excess when uncertainties in the DM
distribution in the Milky Way are accounted for.}\footnote{Note that this
statement does not depend on the annihilation channel or the DM mass, since we
are comparing predicted and measured intensities at the peak of the GeV excess
at 2 GeV, which have to be very similar for \emph{any} DM interpretation of the
\textit{Fermi}  GeV excess.}

\medskip

The situation changes drastically however if current limits would increase by
only a factor of three. This is demonstrated by the blue band in
Fig.~\ref{fig:profile}, which shows the corresponding signal profiles assuming
that $\sv = 0.5\times10^{-26}\cm^3\s^{-1}$.  The allowed signal slope becomes
much steeper since smaller cross-sections require larger DM densities towards
the GC.  \emph{We find that there would be significant tension between measured
and observationally allowed signal morphologies, both towards the Galactic
center ($\psi\lesssim5^\circ$), but even more importantly in the
higher-latitude tail (above $\psi\gtrsim5^\circ$).}

To enforce consistency between the measured and gravitationally allowed signal
morphologies even when dSph limits further strengthen in the future, one would
have to resort to more drastic assumptions, such as a DM profile that
considerably flattens within the inner 1\kpc\ or so, or substructure
enhancement at larger distances from the GC (which however already seems
unlikely to be relevant since due to tidal disruption this effect is expected
to be small at $\psi\lesssim25^\circ$~\cite{Pieri:2009je}), or more complex
theories of particle DM~\cite{Hardy:2014dea}.  We leave the exploration of
these scenarios to future work.

\medskip

Finally, we will estimate the general uncertainty of the annihilation
cross-section that is quoted in Tab.~\ref{tab:fitResults}.  To this end, we
again make use of the above set of valid DM halo models, compatible with the
constraints discussed in Ref.~\cite{Iocco:2011jz} at 95\% CL.  Using all these
models, with the additional constraint that $\gamma\geq1.1$, we find that the
line-of-sight integral over the DM density in direction $\psi=5^\circ$ can vary
by a factor $5.9$ up and by a factor $0.19$ down w.r.t.~the value that is
obtained for our reference profile ($\gamma=1.26$, $r_s=20\rm\;kpc$,
$\rho_\odot=0.4\rm\;GeV\;cm^{-3}$).  Hence, we attribute a generous
astrophysical uncertainty to our best-fit annihilation cross-sections that is
multiplicative and in the range $\mathcal{A} = [0.17, 5.3]$.  Note that some of
the halo models would at face value be too steep in the inner kpc to be
compatible with the \textit{Fermi}  GeV excess morphology; in these cases we
assume that the profile flattens out towards the center.

\section{Conclusions}
\label{sec:conclusions}

In this paper, we presented a critical reassessment of DM interpretations of
the \textit{Fermi}  GeV excess in light of the foreground/background
uncertainties.  To this end, we made use of the results from
Ref.~\cite{Calore:2014xka}, where the emission from the inner $\sim1$ kpc
around the GC as seen at higher latitudes ($|b|>2^\circ$) was robustly
characterized with respect to foreground/background uncertainties.  In
Sec.~\ref{sec:bkg}, we showed that at the peak energy of 2 GeV, all previous
studies of the \Fermi~GeV excess (including the preliminary results from the
\textit{Fermi}-LAT team~\cite{simonaTalk}) suggest a signal morphology that is
consistent with a contracted DM profile.  We entertained here the exciting and
suggestive possibility that all of this emission is due to a signal from DM
annihilation.  Given the complexity of the Galactic foregrounds/backgrounds, we
found that a much larger number of DM models fits the gamma-ray data than was
previously noted, which should be taken into account in future DM searches.  In
particular the low- and high- \emph{tails} of the excess energy spectrum, which
are highly relevant to constrain different DM scenarios, are subject to large
uncertainties.

Our main findings are as follows.
\begin{itemize}
    \item[(i)] We confirmed previous findings that annihilation to gluons, and quark
        final states $\bar{q}q$, $\bar{c}c$ and $\bar{b}b$, provide a good fit.
        However, we found that somewhat higher masses are preferred compared to
        previous analyses, which is due to the additional uncertainty in the
        high-energy tail of the energy spectrum.  In the case of DM
        annihilation into $\bar{b}b$, we found that DM masses up to $m_\chi
        \simeq 74\rm\;GeV$ are allowed (giving a $p$-value above 0.05).
    \item[(ii)] Pure annihilation to pairs of $W$ and $Z$ gauge bosons are excluded
        at a little over $95\%$~CL.  However -- and perhaps surprisingly --
        annihilation to pairs of on shell Higgs bosons produce a rather good
        fit.  Associated gamma-ray lines from $h\to\gamma\gamma$ are close to
        the sensitivity of current instruments.
    \item[(iii)] For annihilation into $\mu^+\mu^-$, the ICS emission plays an
        important role.  We showed that it is possible to find CR propagation
        models for which fits with the ICS emission from $\mu^+\mu^-$ final
        states and $m_\chi \sim 60$--$70$~GeV can become competitive with those
        of annihilation to $\bar{b}b$.  Furthermore, we demonstrated that for
        some models the morphological properties of the GeV excess are well
        reproduced, and thus that it is not possible to exclude ICS emission
        from muons only on the basis of gamma-ray data at \mbox{$|b|>2^\circ$}.
    \item[(iv)] In a realistic model, DM will likely annihilate into a variety of
        channels with different branching ratios.  We remained agnostic to the
        composition of allowed final states and derived constraints on
        different final state triples.  In the case of $\bar{b}b$, $\bar{c}c$ and $\tau^+ \tau^-$
        final states, we found that the best-fit point lies close to the ratios
        predicted for 2HDM, $\bar{b}b:\bar{c}c:\tau^+ \tau^-=0.87:0.08:0.05$.
        We also found that $\mathrm{BR}(\tau^+\tau^-)$ can be substantial (up
        to around 75\%) while still providing a good spectral fit.  In the case
        of annihilation into only $\mu^+\mu^-$ and $\tau^+ \tau^-$, we saw that
        a good fit can always be obtained when
        $\mathrm{BR}(\mu^+\mu^-)\gtrsim0.6$ and $m_{\chi}\sim50$~GeV.  When
        allowing for additional annihilation into $\bar{q}q$ or $\bar{b}b$, we
        found that all values of $\mathrm{BR}(\mu^+\mu^-)$ result in a good
        fit.  Lastly, for the annihilation into $hh$, we found that a good fit
        is obtained only when $\mathrm{BR}(hh)\gtrsim0.8$ and when the DM mass
        is close to $m_h$.
    \item[(v)] For hidden sector models with a light mediator $\phi$ that subsequently
        decays into combinations of $e^{+}e^{-}$, $\mu^{+}\mu^{-}$ and $\pi^{+}\pi^{-}$
        -- much like in the case of direct annihilation to $\mu^{+}\mu^{-}$ -- the ICS emission dominates. 
        We found that the allowed DM particle mass range is between 45
        and  125 GeV, with cross-sections that are constrained (but not severely) 
        by the AMS positron fraction data.
    \item[(vi)] We showed that, given dynamical and microlensing constraints on the DM
        halo, \emph{current limits from dwarf Spheroidal galaxies} are not yet
        in tension with the DM interpretation of the \textit{Fermi}  GeV
        excess.  However, we also demonstrated that if these limits further
        strengthen by a factor of three, there would be significant tension
        between measured and observationally allowed \emph{morphologies} of the
        GC signal.  We furthermore showed that given current constraints, the
        annihilation cross-section is uncertain by a factor of about five up
        and down (at 95\%~CL). 
\end{itemize}

\medskip

\noindent
The covariance matrix as well as the flux of the \textit{Fermi}  GeV excess
emission that are used for the spectral analysis are available
online.\footnote{\url{https://staff.fnwi.uva.nl/c.weniger/pages/material/}}

\begin{acknowledgements}
    We thank Markus Ackermann, Carmelo Evoli, Dan Hooper, Gudlaugur
    J$\acute{\textrm{o}}$hannesson, Simona Murgia, Troy Porter and Neal Weiner for useful
    discussions.  The research of C.~W. is part of the VIDI research programme
    ``Probing the Genesis of Dark Matter'', which is financed by the
    Netherlands Organisation for Scientific Research (NWO).  This work has been
    supported by the US Department of Energy. F.~C. and C.~M.
    acknowledge support from the European Research Council through the 
    ERC starting grant
    WIMPs Kairos, P.I. G. Bertone.

    \bigskip
    \paragraph*{Note added.} While finalizing this work, a preprint
    appeared~\cite{Agrawal:2014oha} that also considers quarks and massive
    gauge boson final states in light of the background model systematics from
    Ref.~\cite{Calore:2014xka}.  For the cases where our analyses overlap we
    find agreeing results.
\end{acknowledgements}

\bibliography{GCfit.bib}
\bibliographystyle{JHEP}

\end{document}